\documentclass[
 amsmath,amssymb,
aps,pre,twocolumn,
]{revtex4-2}
\usepackage{amsmath}
\usepackage{graphicx}
\usepackage{xcolor}
\usepackage{dcolumn}
\usepackage{bm}
\usepackage{ulem}
\usepackage[page]{appendix}
\usepackage{cases}  
\usepackage{upgreek}
\definecolor{colortodo}{RGB}{255,0,0}

\definecolor{colortodo2}{RGB}{0,0,255}

\begin{document}

\title{Dynamical Boundary Following and Corner Trapping of Undulating Worms}

\author{Sohum Kapadia}
\author{Arshad Kudrolli}
\affiliation{Department of Physics, Clark University, Worcester, MA 01610}

\date{\today}
\begin{abstract}
We investigate the behavior of {\it Lumbriculus variegatus} in circular and polygonal chambers and show that the worms align with the boundaries as they move forward and then become dynamically trapped at the concave corners over prolonged periods. We model the worm as a self-propelled rod and derive analytical expressions for the evolution of its orientation when it encounters the flat and the circular boundaries of the chamber. By further incorporating translational and rotational diffusion, arising due to the undulatory and peristaltic body strokes, we demonstrate through numerical simulations that the self-propelled rod model can capture both the boundary aligning and the corner trapping behavior of the worm. The Péclet number $Pe$, representing the ratio of forward propulsion to rotational diffusion, is found to characterize the boundary alignment dynamics and trapping time distribution of the worm. Simulations show that the angle of the worm's body with the boundary while entering a concave corner plays a key role in determining the trapping time, with shallow angles leading to faster escapes. Our study demonstrates that directed motion combined with limited angular diffusion can lead to spatial localization that mimics shelter seeking behavior in slender undulating limbless worms, even in the absence of thigmotaxis or contact seeking behavior.
\end{abstract}

\maketitle

\section{Introduction}
The tendency of living organisms and synthetic agents to follow boundaries and accumulate in corners is a common behavior observed across a wide range of active matter systems. From individual motile cells to large animals and even robotic swarms, these behaviors play an essential role in navigation, aggregation, and spatial organization. For example, sperm cells rely on contact with boundaries to navigate the convoluted structure of the reproductive tract and reach the egg~\cite{nosrati2016predominance,denissenko2012human,guidobaldi2014geometrical}. In larger animals, such as flies, rats, cockroaches, fishes, and humans, contact-driven behavior, referred to as thigmotaxis, is often used to locate shelter and navigate disordered environments~\cite{gotz1985centrophobism,besson2005centrophobism,creed1990interpreting,sharma2009function,kallai2007cognitive,walz2016human,Zhang2023}. These behaviors are not limited to biological systems. Collections of mechanical robots have also been shown to spontaneously align along boundaries and accumulate in corners, resembling the behavior of infant rats placed in enclosed spaces~\cite{May2006ratpups}.

In active matter systems composed of elongated polar particles, such as synthetic rods or hex-bots, steric interactions can prevent rotation, leading to collective trapping near boundaries and corners ~\cite{kudrolli2008swarming,Kaiser2013Capturing,Kumar2019Trapping,Brown2024Boundary,Giomi2013Swarming,Deblais2018Boundaries}. At the microscopic scale, organisms like {\it Escherichia coli}  and micro-algae like \textit{Chlamydomonas reinhardtii} are known to accumulate at surfaces and align with them~\cite{frymier1995three,Lauga2006Swimming,berke2008hydrodynamic,Li2009microswimmers,Lauga2009,Jung2010,kantsler2013ciliary,yuan2015hydrodynamic,Sipos2015Hydrodynamic,Ostapenko2018Curvature,Ohmura2018ciliamechanosense}. Although the behavior appears similar across scales and systems, the mechanisms responsible can differ widely due to variations in physiology and the modes of locomotion.

In micron-scale systems such as sperm and bacteria moving in fluids, the dynamics occur in a low Reynolds number regime where viscous forces dominate and inertia is negligible. Under these conditions, long-range hydrodynamic interactions with nearby surfaces can cause alignment and boundary following, and there is in fact no direct contact with the boundary~\cite{Lauga2009}. Other studies have proposed that direct mechanical interactions, including surface contact by cilia, can generate torques that reorient organisms with respect to nearby surfaces~\cite{Li2009microswimmers,kantsler2013ciliary}. Computational models have also shown that active particles can travel long distances along boundaries by responding to spatial variations in surface geometry, a behavior known as topotaxis~\cite{Schakenraad2020,Sadjadi2024}. However, clear experimental evidence that tracks full-body motion and reveals the specific mechanisms at play in actual organisms is still limited.

Despite the prevalence of boundary and corner aggregation across diverse systems, the underlying mechanisms often remain ambiguous. In particular, few studies have examined situations where interactions with boundaries rely solely on the sense of touch, without any guidance from light, chemical gradients, gravity, or from interactions with other organisms. Yet this sensory condition is quite common in nature, and is even experienced by humans navigating in complete darkness. Understanding how organisms and particles behave in such tactile-only environments may offer new insights into the physics of navigation, collective motion, and spatial organization in noisy active systems.

\begin{figure*}
 \includegraphics[width = 0.65\textwidth]{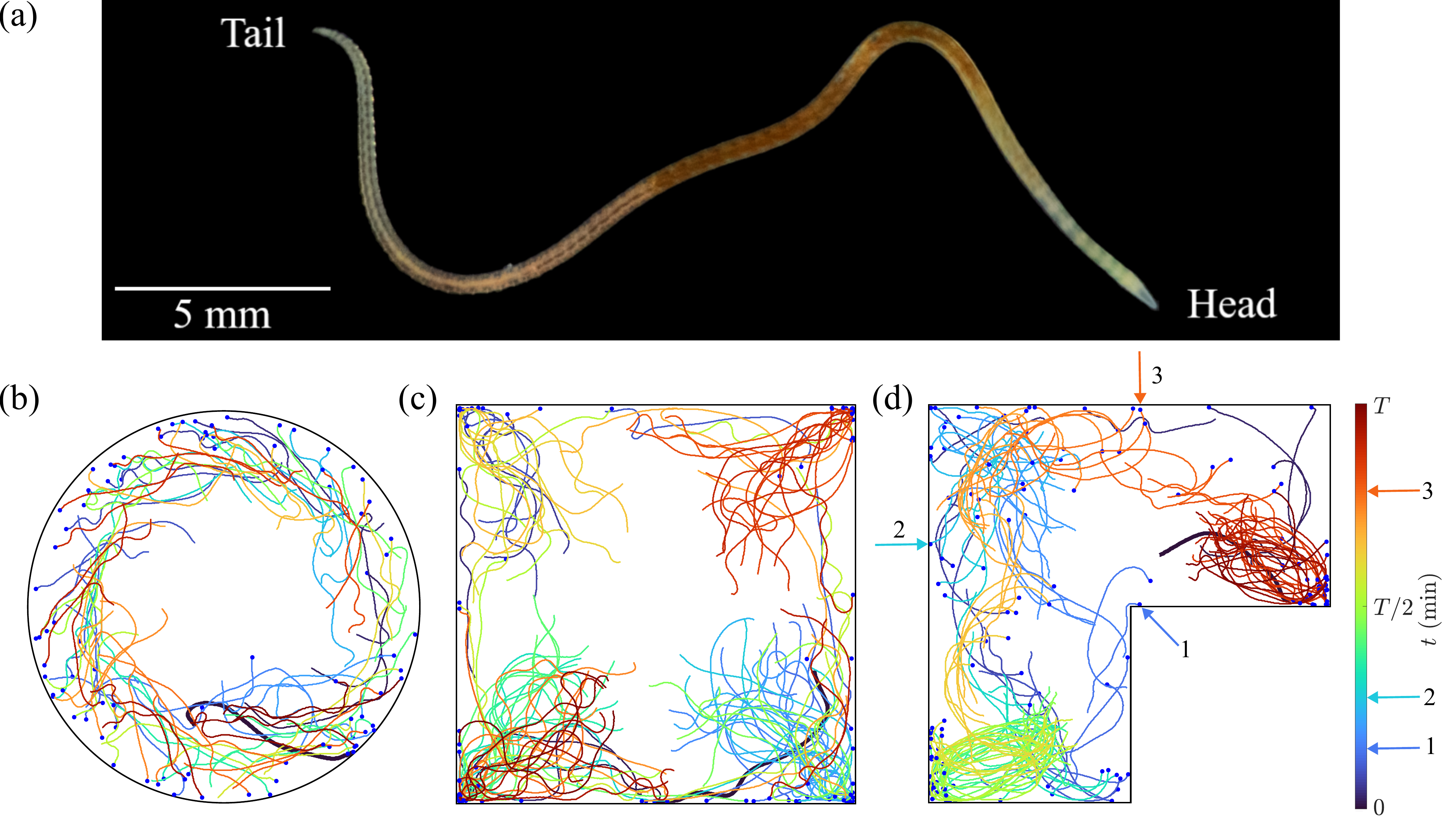}
  \caption{(a) An image of \textit{Lumbriculus variegatus},  also known as  California blackworm being studied. (b-d) Superposed snapshots of the worm at $\Delta t = 10$\,seconds time intervals moving inside a circular chamber with diameter $d = 4$\,cm (b), a square chamber with side length $l= 4 \textrm{ cm}$ (c), and a polygon chamber with long sides $l = 4 \textrm{ cm}$ (d),  over total observation time $T = 18$\,min, 21\,min, and 25\,min, respectively. The snapshots correspond to time $t$ denoted by the color map. For reference, the snapshot at $t = 0$ is highlighted in bold. The location of the head is indicated by a solid blue marker and is observed to stay close to the boundary in the circular chamber, and predominantly near the concave corners in the square chamber. The head remains close to the concave corner and does not follow the boundary around the convex corner in the polygonal chamber. The arrows labeled 1, 2, 3 in the color bar indicate three spontaneous departures of the worm from the polygonal chamber boundary in (d).}
 \label{fig:convex_corner}
\end{figure*}

\textit{Lumbriculus variegatus}, also known as the California blackworm, belonging to the Oligochaeta subclass of the Annelida phylum are found widely in the sediments at the bottom of shallow fresh water bodies.  \textit{L. variegatus}, along with \textit{T. tubifex}, have become an important model to study active matter in the laboratory~\cite{drewes1990giant,lesiuk1999autotomy,drewes1990morphallaxis,drewes1999worm,drewes1999helical,drewes1989hindsight,kudrolli2019burrowing,biswas2023,patil2023ultrafast,nguyen2021emergent,deblais2020phase,deblais2020rheology,ozkan2021collective,zirbes2012self,deblais2020rheology,heeremans2022chromatographic}. These centimeter-scale slender worms typically move in water and soft sediments with transverse undulatory and peristaltic strokes~\cite{kudrolli2019burrowing}, they can disperse rapidly over short distances with helical strokes or body reversals when threatened~\cite{drewes1999helical,patil2023ultrafast}, and show a rich set of behavior including forming blobs that can adaptively move around in response to external stimuli such as temperature, light and physical stresses~\cite{nguyen2021emergent,ozkan2021collective,zirbes2012self,deblais2020rheology}.  Recently, it has been shown that \textit{L. variegatus} uses its prostomial nerves to probe physical obstacles and follows the boundary while moving in a circular chamber, even when occasionally switching directions~\cite{biswas2023}. This boundary following behavior could emerge either due to positive thigmotaxis  to seek out and remain in contact with surfaces  or simply due to its dynamics in a forward direction while its head is in contact with the boundary. The dominant mechanism responsible, whether caused by an active attraction to touch (thigmotaxis) or is a passive result of its movement when touching the boundary, could not be distinguished in the previous study, which only investigated concave chambers with constant curvatures.

In this article, we examine the spatial distribution of a single worm confined to water filled chambers, both with and without concave and convex corners. Transparent chambers are constructed to allow full body tracking of the aquatic worm, enabling the investigation of how boundary curvature influences behavior near impenetrable walls. Our observations reveal that the worm aligns progressively with the boundary upon contact and eventually becomes trapped in a concave corner of the chamber. In contrast, the worm continues into the interior of the chamber without following the boundary at the convex corner. Using both square and circular chambers, we study the travel distance over which the worm aligns with the boundary and the time scales associated with trapping at the corners. These observations allow us to quantitatively investigate the physical mechanisms underlying the observed localization behavior, supported by complementary simulations that model the worm as a self propelled rod. The simulations reproduce many of the key features observed in experiments, demonstrating that boundary following and localization at corners can arise purely from the passive dynamics of long slender organisms when their head comes in contact with the boundary.

\section{Experimental Methods}

An image of a California blackworm used in our study is shown in Fig.~\ref{fig:convex_corner}(a). All the worms belonged to the same stock and were maintained using the protocols described in Ref.~\cite{biswas2023}. A worm with length $l_w = 20 \pm 3$ mm is picked at random from a reservoir and placed in a quasi-two dimensional observation chamber. The chamber is constructed by laser-cutting holes in clear acrylic sheets, and sandwiching them between clear acrylic sheets in a larger water-filled container. The thickness of the resulting chamber $h=1.5$\,mm is chosen to be much larger than the worm diameter $d_w \approx 100 \  \upmu\textrm{m}$, following prior protocol \cite{biswas2023}.  

The system is imaged from above with a digital camera over approximately $30$\,minutes at 24\,frames per second (fps) while the worm is back-lit. The projected shape of the worm in the horizontal plane is obtained through image processing. In order to average over variations in individual worm size and behavior, the experiments are repeated with different worms for about $n_T = 10$ trails. Further information on the imaging and full body shape tracking techniques can be found in the Supplementary Information (SI)~\cite{sup-doc}.   

\section{Emergent Dynamics}
\subsection{Effect of chamber shape}
\label{sec:boundaryinter}
Figure~\ref{fig:convex_corner}(b-d) shows three representative chambers with increasingly complex shapes used in our investigations along with snapshots of the worm, illustrating the typical dynamics observed over time interval $T$. The corresponding Movies S1, S2, and S3 can be found in the  SI~\cite{sup-doc}. In the circular chamber shown in Fig.~\ref{fig:convex_corner}(b), we observe that the worm essentially follows the boundary with the head close to the boundaries as in a previous study~\cite{biswas2023}. Whereas, the worm is not only found near the boundaries, but also spends significant time in the concave corners of the square chamber and polygonal chamber as shown in Fig.~\ref{fig:convex_corner}(c) and  Fig.~\ref{fig:convex_corner}(d), respectively. In all cases, the body undulations can be noted to cause significant excursions of the worm body which lead it to have only intermittent direct contact with the boundary. Thus, while the body interacts frequently with the boundary, the contacts can be noted to be collisional and noisy. This makes the interactions of the California blackworm with the boundaries qualitatively different from microorganisms which experience a long range interaction with boundaries~\cite{Lauga2009,Li2009microswimmers}. To specifically understand the hydrodynamic regime of the system we estimate the Reynold's number $Re = \rho v_h l_w /\mu \approx 40$, where $\rho = 10^{3}$ kg m$^{-3}$ is the density of water, $v_h \approx 2$ mm s$^{-1}$ is the average speed of the worm's head, $l_w \approx 20$ mm is the average worm length and $\mu = 10^{-3}$ Pa s is the dynamic viscosity of water. At the intermediate Reynold's number, the flow field around the worm decays as $1/r^3$, where $r$ is the separation between the worm and the chamber. The worm diameter is $d\approx 0.1$ mm, and the chamber gap height is $h = 1.5$ mm, so the radial distance is of order $r \approx 10 \ d$, at such far distance, the hydrodynamic interactions are negligible.

Figure~\ref{fig:convex_corner}(d) also illustrates that the worm does not curve round the convex corner in the polygon chamber to follow the boundary, rather it looses contact with the boundary while continuing to move in the interior region of the chamber but rather continues into the central portion of the chamber. Moreover, we consistently observe this behavior across multiple trials of the experiment. It can be also noted in Fig~\ref{fig:convex_corner}(d) that the worm departs the flat boundaries at time $t = 4$\,min, $7$\,min and $19$\,min. These observations lead us to conclude that the worm does not show any specific attraction to the boundary. Combined with the fact that the worm appear to leave the relative shelter of the corners suggests that thigmotaxis is not the reason why the worms are found near the boundary and the corners of the chamber. Rather, these observations suggests a dynamical mechanism for the aggregation observed near the boundary and the concave corners. 

\begin{figure*}[htbp!]
  \centering
  \includegraphics[width = 0.7\textwidth]{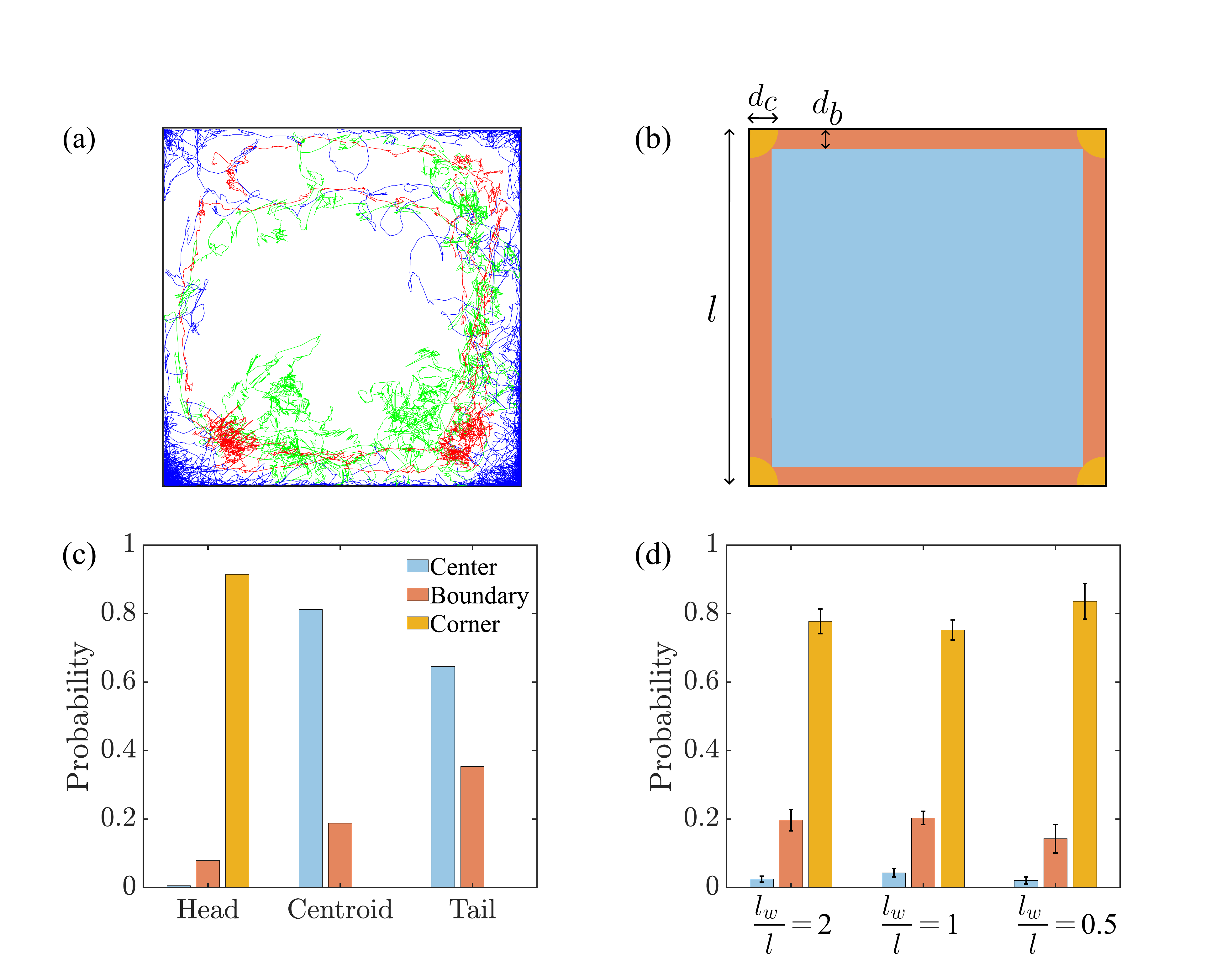}
   \caption{(a) The trajectory of the head (blue), centroid (red), and tail (green) of a worm moving in a square chamber over time $T = 10$\,minutes  ($l_w = 23$\,mm; \textit{l} = 4\,cm). (b) The corner, boundary, and center regions in the chamber. (c) The probability of finding the head, centroid, and tail in various parts of the chamber shows the localization of the head near the corners. (d) The probability of finding the head in square chambers with worm to side length aspect ratio, $l_w/l$ = 2, 1 and 0.5 all show localization of the head at the corners ($l_w  = 20 \pm 3$\,mm).  The error bars denote root mean square deviations of a trial from the mean ($n_T =10$~trails; $T=30$~minutes each).}
  \label{fig:spatial}
 \end{figure*} 
 
\subsection{Spatial distribution}
The tracked positions of the worm's head $r_h$, tail $r_t$, and centroid $r_c$ moving in a square chamber with side length $l = 4$\,cm is plotted in Fig.~\ref{fig:spatial}(a). The spatial position of the worm in the chamber is divided into three regions to quantify their distributions as shown in Fig.~\ref{fig:spatial}(b). The corner regions correspond to distances less than $d_c$ from each corner. The boundary regions correspond to distances that are less than  $d_b = \frac{d_c}{\sqrt{2}}$ from the boundaries, excluding those which belong to the corner regions. The central region  correspond to the complementary locations that do not belong to either the corner or the boundary regions. The distance $d_c$ is selected to ensure that movement involving a brief loss of contact (characterized by short stroke durations) are still spatially categorized as instances of interaction with the corner or boundary regions. To quantitatively characterize the fluctuations caused by the body strokes, we calculate the root mean square (RMS) displacement defined as $\sqrt{\langle\Delta r^2\rangle} = \sqrt{\frac{1}{N}{\sum_{i = 1}^{N}|r(t_i+\Delta t) -r(t_i)|^2 }}\,,$ where $\langle .. \rangle$ denotes averaging over time, $r(t_i)$ represents $r_h$, $r_t$, and $r_c$ at time $t_i$, $i$ is the timestep index, $\Delta t$ is the time interval, and $N$ is the total time steps. We choose $\Delta t = 2.5$\,s, corresponding to twice the peristaltic stroke period to average out the intra-stroke fluctuations. Over this time interval and the RMS displacement values of the head and centroid positions, while away from the boundaries, are $\sqrt{\langle\Delta r_h^2\rangle} = 5.74 \pm 0.74 $\,mm and $\sqrt{\langle\Delta r_c^2\rangle} = 2.64 \pm 0.63 $\,mm, respectively. Whereas,  $\sqrt{\langle\Delta r_h^2\rangle} = 1.14 \pm 0.07 $\,mm, and $\sqrt{\langle\Delta r_c^2\rangle} = 1.36 \pm 0.20 $\,mm while the worm is near a corner, due to the interactions with the boundary. The mean and standard deviation of the RMS displacement values are calculated across total of $n_T = 10$ trials, each performed over $n_T = 30$ minutes. Accordingly, we choose $d_c = 3$\,mm to account for most of the short time departures from the corner which arise due to fast body strokes.

The probability of finding the head, the centroid and the tail in the different regions normalized by respective area are shown in Fig.~\ref{fig:spatial}(c) for the trajectory in Fig.~\ref{fig:spatial}(a). We observe that the head is found in the corners over 88\% of the time in this case, whereas, the tail and the centroid are typically found in the central or boundary regions.  
Figure~\ref{fig:spatial}(d) shows the probability of finding the head in the different regions for chambers with side length, $l=10$\,mm, $20$\,mm and $40$\,mm for $n_T = 9$, $10$ and 10 trials, each performed over $T \simeq 30$\,mins, respectively. We characterize the system using an aspect ratio defined as ratio between the worm length and chamber length. In Fig. \ref{fig:spatial}(d), corresponding to $l=10$\,mm, $20$\,mm and $40$\,mm, $l_w/l = 2, 1$ and 0.5 for worm length $l_w = 20\pm 3$ mm, respectively. Across all three experimental configurations, the worm's head was consistently observed to reside near the chamber corners. This indicates the robustness of concave corner trapping persists even when the worm is strongly confined ($l_w/l =2$) and its length exceeds the chamber side length.

 \begin{figure*}[htbp!]
\centering
\includegraphics[width = 0.7\textwidth]{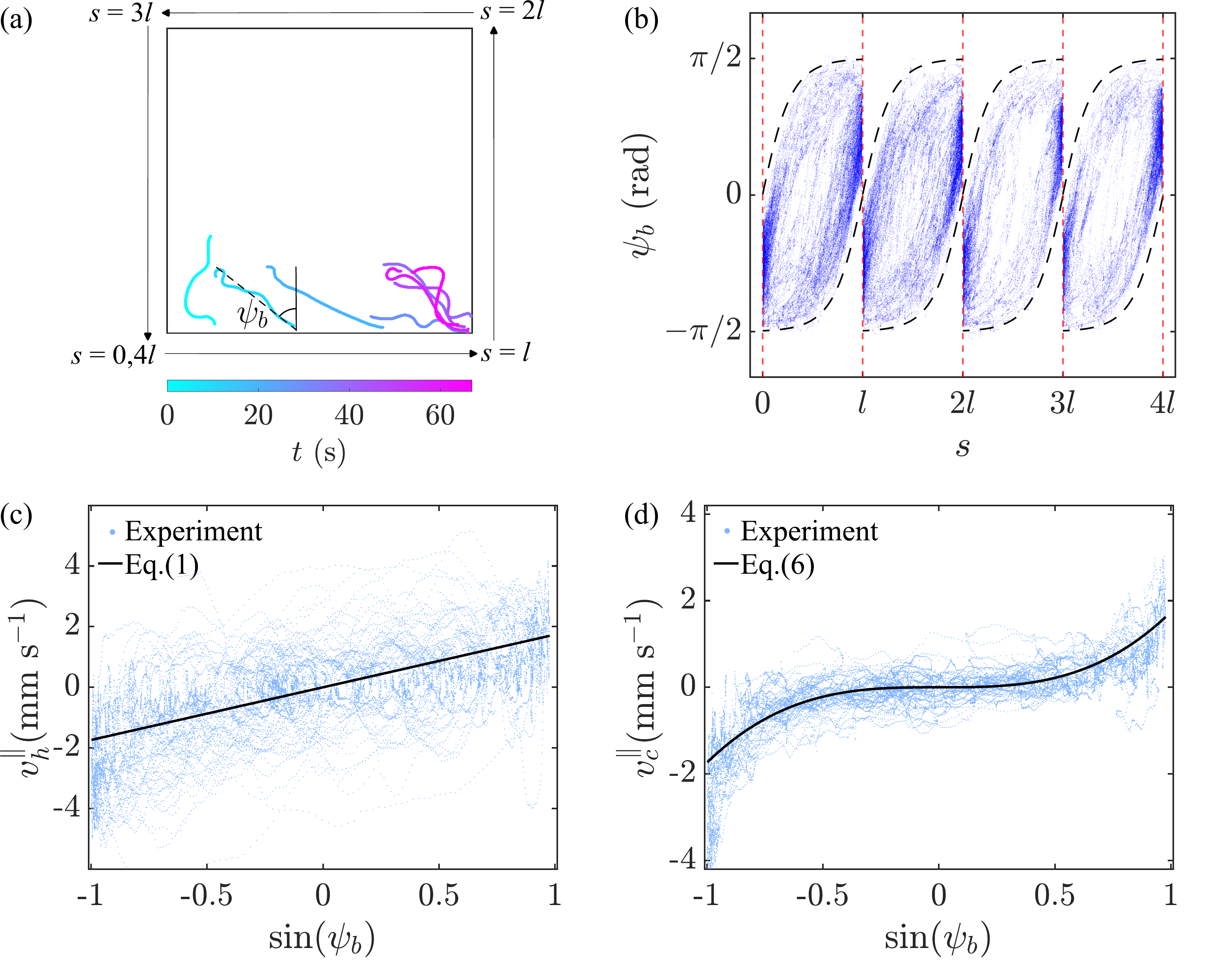}
\caption{(a) Snapshots of a worm as it aligns with a boundary and gets trapped in a corner. The time corresponds to the color map.  (b) A scatter plot of body orientation $\psi_{b}$ versus distance along perimeter $s$. The  dashed (black) lines correspond to Eq.~(\ref{eq:psib}) calculated with the kinematic model, and the vertical dotted (red) lines correspond to the worms trapped at the corners.  (c) The scatter plot of $v_h^{||}$ as a function of $\sin\psi_b$ recorded each time a worm contacts a boundary, shows increasing trend which is well captured by a linear fit (solid black line) based on Eq.~(\ref{eq:linear_eqn}) with $v_h = 1.74$\,mm\,s$^{-1}$. (d) The scatter plot of the measured worm centroid velocity $v_c^{||}$ as a function of $\sin(\psi_b)$ is described by Eq.~(\ref{eq:v_c}) assuming $v_r$ is the same as $v_h$ obtained by fitting Eq.~(\ref{eq:linear_eqn}).} 
\label{fig:boundary}
\end{figure*}

\subsection{Boundary aligning  dynamics}
We characterize the behavior of the worm while its head is located in the boundary region of the chamber. Figure~\ref{fig:boundary}(a) shows illustrative snapshots of a worm as it moves forward and increasingly aligns with  the chamber boundary before arriving at a corner. While trapped there, the worm usually reorients its body along the corner angle bisector as it continues to perform undulatory and peristaltic strokes. 
The orientation of the worm when it encounters the boundary is quantified by the angle $\psi_b$ that the line joining the head and the centroid of the worm makes with the normal to the boundary (see Fig.~\ref{fig:boundary}(a)). In Fig.~\ref{fig:boundary}(b), we plot $\psi_b$ as a function of the head position $s$ along the chamber boundary, measured counterclockwise from the bottom-left vertex. The data correspond to instances when the head is in contact with the boundary, collected over $n_T = 10$ independent trials, each of duration $T = 30$\,minutes. The strokes induce time-dependent deformations of the worm's body, leading to fluctuations in both its interactions with the boundaries and the direction of motion. Despite this variability, the data points are not uniformly scattered, but instead show well defined trends. 

The boundary-aligning behavior of the worm illustrated in Fig.~\ref{fig:boundary}(a) gives rise to increasing magnitude of $\psi_b$ as the head position $s$ varies from $0$ to $l$ in Fig.~\ref{fig:boundary}(b).  The scatter plot reveals that the worms are rarely oriented perpendicular to the boundary midway between adjacent corners. Because of the four fold rotational symmetry of the chamber, the distributions repeat after each $l$. Further, the points above and below $\psi_b=0$\,rad denote counterclockwise and clockwise motion along the four sides of the chamber, respectively, and are thus related to each other via a $\pi$-rotation. 
In addition to the points which fall on the curves corresponding to the worm moving along the boundary, a large number of points can be found near the dotted vertical lines around $s = 0$, $l$, $2l$, $3l$, and  $4l$ in Fig.~\ref{fig:boundary}(b). When the worm gets trapped near the corner, $\psi_b$ fluctuates between  $\psi_b = 0$ and $\pi/2$\,rad, or  $\psi_b = -\pi/2$\,rad and $0$, depending on whether the worm is moving anticlockwise or clockwise, respectively. Thus, the observed clustering of points in the scatter plot represents boundary aligning and trapping behavior along each of the chamber sides, and corners. 

\begin{figure*}[htbp!]
\centering
\includegraphics[width =  0.7\textwidth]{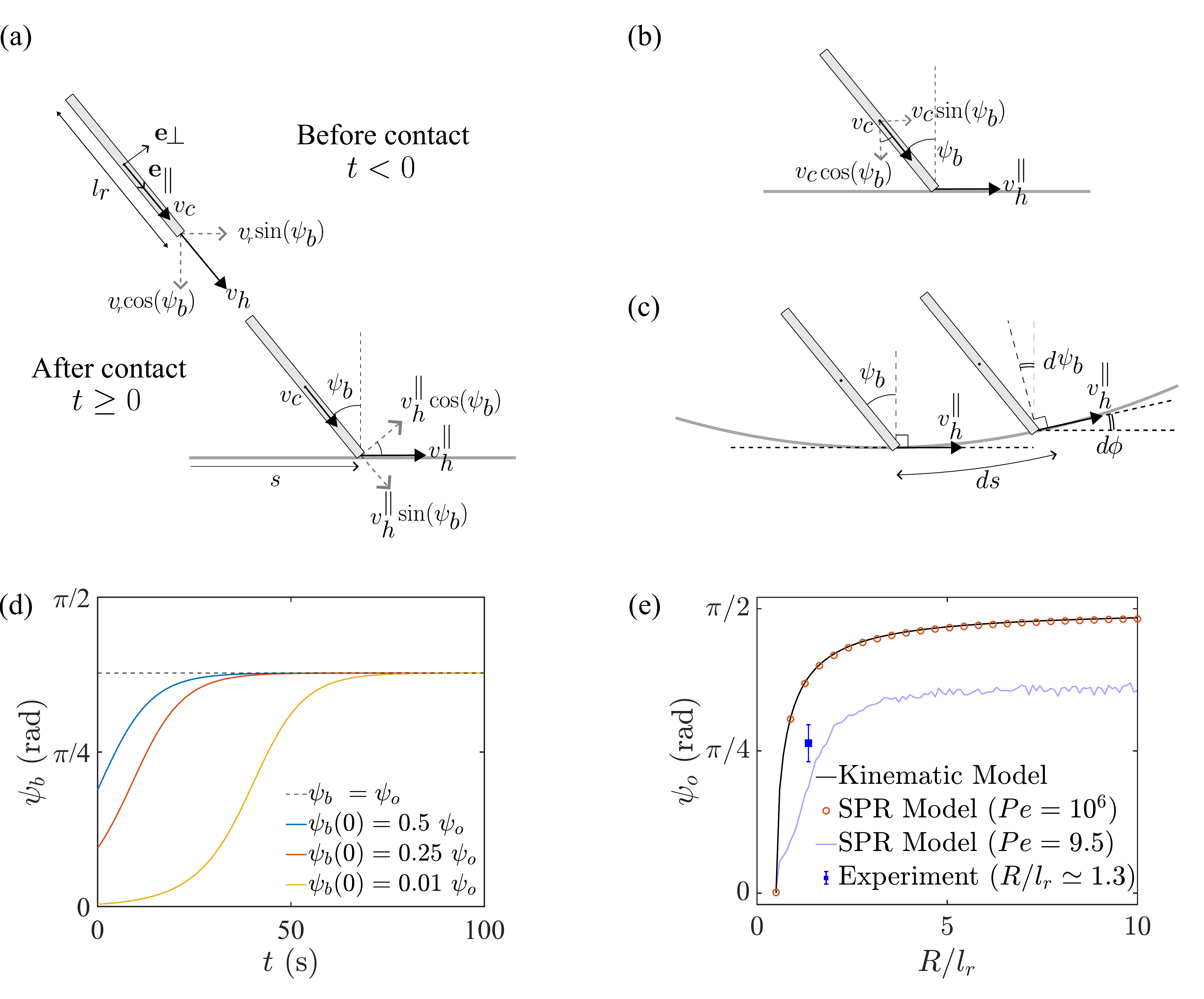}
\caption{(a-c) Schematics of a self propelled rigid rod of length $l_r$ interacting with the boundary. The unit vectors $\mathbf{e}_\parallel$ and $\mathbf{e}_\perp$ denote directions parallel and perpendicular to the rod axis, respectively. $v_c$ and $v_h$ are the velocities of the rod center and head, respectively. (a) The rod head velocity components  before and after contact with a flat boundary. (b) The velocity components of the rod center parallel and perpendicular to the flat boundary. (c) A rod on a curved boundary, illustrating the variation $d\psi_b$ in its orientation resulting from a parallel displacement by distance $ds$ along the curve while the slope angle changes by $d\phi$. (d) $\psi_b(t)$ obtained by numerical integration of Eq.~(\ref{eq:psit}) for three different initial conditions ($R = 20$\,mm; $l_r=15$\,mm). Each example converges to the stable value $\psi_o$ consistent with Eq.~(\ref{eq:Stable angle}). (e) The evolution of the stable orientation of a rigid rod moving in a circular chamber as a function of $R/l_r$. }
\label{fig:analytic soln}
\end{figure*}

We further characterize the boundary interaction by plotting the projection of the head velocity along the boundary $v_h^\parallel$ as a function of $\sin(\psi_b$) in Fig.~\ref{fig:boundary}(c). While the data shows considerable scatter, the average trend can be described by,
 \begin{equation}
     \label{eq:linear_eqn}
      v_h^\parallel = v_{h} \sin{\psi_b},
 \end{equation}
where $v_h$ is the average speed of the worm's head, while it is away from the boundary. The fit (solid black line in  Fig.~\ref{fig:boundary}(c)) to the data yields $v_h = 1.74$\,mm\,s$^{-1}$, as compared with $v_h = 1.72 \pm 0.65$\,mm\,s$^{-1}$ measured in the experiments (see SI document).  Hence, the worm essentially collides inelastically with the boundary, losing most of its momentum perpendicular to the wall, while preserving the parallel component, indicative of a sliding contact. 

In the following, we model the worm dynamics based on the observation that it moves along the mean orientation of its body, on average and interacts with the boundary through steric repulsive forces, resulting in frictionless sliding of the head along the boundary, as described by Eq.~(\ref{eq:linear_eqn}).

\section{Self-propelled rod analysis}
\subsection{Kinematic model of boundary alignment} 
When a self-propelled worm-like organism encounters an immovable boundary, its motion becomes constrained by the geometry of the environment. Rather than continuing along its original trajectory, the organism slides along the boundary while simultaneously reorienting. This reorientation arises because the steric repulsive force exerted by the boundary acts at the head, rather than through the center of mass, generating a torque that rotates the body. The magnitude of this rotation depends on the lever arm and the component of the head’s velocity perpendicular to the rod axis. We formalize this mechanism by decomposing the sliding velocity along the boundary into parallel and perpendicular components relative to the rod axis and relating these to the rate of change of the orientation angle.

Hence, we simplify our analysis of worm kinematics by approximating the worm as a self-propelled rod (SPR) of fixed length $l_r$, corresponding to the worm’s average head-to-tail distance $l_e$. Schematics of the rod before and after it contacts a flat boundary at time $t = 0$ are shown in Fig.~\ref{fig:analytic soln}(a).  Unit vectors $\mathbf{e}_\parallel$ and $\mathbf{e}_\perp$ denote directions parallel and perpendicular to the rod's axis, respectively. 
Prior to contact, we assume that the rod propels with constant speed $v_r$ along the unit vector $\mathbf{e}_\parallel$ aligned with the rod axis. As a result, both the rod center velocity $v_c$ 
and head end velocity $v_h$ are equal to $v_r$ as the rod approaches the boundary. 

When the rod encounters the boundary,  a reaction force acts at the point of contact which constrains the rod to slide along the boundary with velocity $v_h^\parallel = v_r \sin(\psi_b)$, given by Eq.~(\ref{eq:linear_eqn}) for the worms. Then, $v_h^\parallel$ can be decomposed into parallel and perpendicular components to the rod axis, given by $v_h^\parallel \sin(\psi_b)$ and $v_h^\parallel\cos(\psi_b)$,  respectively. 
Since the worm is modeled as a frictionless rigid rod of length $l_r$, and the boundary exerts a steric repulsive force on the head (located at one end of the rod), this force does not act through the rod's center of mass. Instead, it acts at a distance of $l_r/2$ from the center, generating a torque.
This torque leads to an instantaneous rotation of the rod about its center. The rate of change of the orientation angle $\psi_b$ is therefore determined by the perpendicular component of the head’s velocity, which is $v_h^\parallel \cos(\psi_b)$, divided by the lever arm $l_r/2$. This yields the angular velocity:
\begin{equation}
    \frac{d\psi_b}{dt} =  \frac{v_h^\parallel \cos(\psi_b)}{l_r/2}.
    \label{eq:vs}
\end{equation}
Substituting the sliding boundary condition for the rod $v_h^\parallel = v_r \sin(\psi_b)$ into Eq.~(\ref{eq:vs}), and using trigonometric identities, we have 
\begin{equation}
      \frac{d\psi_b}{dt} = \frac{v_r}{l_r} \sin(2\psi_b).   
\end{equation}
Upon integration,  
\begin{equation}
\psi_b(t)  = \tan^{-1}\left(\tan(\psi_b(0))\exp\left(\frac{2v_rt}{l_r}\right)\right)\,,
\end{equation}
where $\psi_b(0)$ is the incident orientation, when the head  comes in contact with the boundary at time $t=0$. Solving for the head position $s$ starting with $\frac{ds}{dt} = v_r \sin(\psi_b)$, and using the chain rule $\frac{ds}{dt} = \frac{ds}{d\psi_b}\frac{d\psi_b}{dt}$, we get
$\frac{ds}{d\psi_b} = \frac{l_r}{2\cos(\psi_b)}$. 
Integrating, 
$ {s - s(0)} = \frac{l_r}{2} \ln \left( \sqrt{\frac{1 + \sin(\psi_b)}{1 - \sin(\psi_b)}} \right) - \frac{l_r}{2} \ln \left( \sqrt{\frac{1 + \sin(\psi_b(0))}{1 - \sin(\psi_b(0))}} \right)$.
Rearranging, 
\begin{equation}
    \psi_b(s) = \sin^{-1}\left(\frac{ K\exp\left( \frac{4(s-s(0))}{l_r} \right) - 1}{ K\exp\left( \frac{4(s-s(0))}{l_r} \right) + 1}\right),
\label{eq:psib}
\end{equation}
where $K = \frac{1 + \sin(\psi_b(0))}{1 - \sin(\psi_b(0))}$ and $s(0)$ is the incident position of the head, along the boundary. Thus, California blackworm can be expected to orient along the boundary over the scale of a body length after first coming in contact with the boundary. 

We plot Eq.~(\ref{eq:psib}) as dashed lines in Fig.~\ref{fig:boundary}(b) for $s(0) = 0$, $l$, $2l$, $3l$ and $4l$, and $\psi_b(0) = 0.01$\,rad, corresponding to counterclockwise motion along each side, and $l_r= l_e \approx 15$\,mm, obtained from the measurements. Using symmetry, we also plot $\psi_b(s)$  while moving clockwise from $4l$ to $3l$, $2l$, $l$ and $0$. The curves in Fig.~\ref{fig:boundary}(c) enclose the scattered points, which is expected as the worm starts near the corners, initially at an angle to the boundary, and then aligns with it.

The speed of the worms centroid along the boundary can be also estimated using this kinematic model. 
The rod center moves instantaneously  along $\mathbf{e}_\parallel$ while the rod rotates at the point of contact (see Fig.~\ref{fig:analytic soln}(a)), so that $v_c = v_h^\parallel \sin(\psi_b)$. Applying the sliding condition at the rod's head, we have $  v_h^\parallel = v_r \sin(\psi_b)$ Then, the component of rod center velocity along the boundary shown in  Fig.~\ref{fig:analytic soln}(b) is given by,
\begin{align}
     \label{eq:v_c}
     v_c^\parallel = v_c \sin(\psi_b) = v_h^\parallel  \sin^2(\psi_b) = v_r  \sin^3(\psi_b).
\end{align}
We plot $v_c^\parallel$ as a function of $\sin(\psi_b)$ in Fig.~\ref{fig:boundary}(d), along with Eq.~(\ref{eq:v_c}) using $v_r = 1.74 \textrm{ mm s}^{-1}$, which was obtained by fitting Eq.~(\ref{eq:linear_eqn}). We observe that the calculated $v_c^\parallel$ captures the main trends in the data, despite additional fluctuations arising from the body strokes that are not accounted for in the kinematic model. 

The kinematic model can also be extended to capture boundary aligning behavior observed in the circular chamber. We consider a curved boundary represented in parametric form as $\phi(s)$, where $s$ denotes the arc length along the curve and $\phi$ is the angle between the tangent and the horizontal.
The evolution of the rod's orientation $\psi_b$, defined relative to the curve's normal is a cumulative effect of the rod's angular velocity and the boundary curvature. To elaborate the effect due to curvature, for an infinitesimal distance $ds$, the change in the slope of the curve is $d\phi$ as shown in Fig.~\ref{fig:analytic soln}(c). To illustrate the decrease in $\psi_b$ by $d\psi_b$ = -$d\phi$, the rod is shifted parallel to its original orientation at the new position for reference. It is seen that an increase in the slope of the curve by $d\phi$ leads to a decrease in $\psi_b$ by $d\psi_b = -d\phi$. Assuming sufficiently slow variation in $\phi(s)$, the two effects add up linearly resulting in a net angular speed given by,  $$\frac{d\psi_b}{dt} = \frac{v_r}{l_r} \sin(2\psi_b) -\frac{d\phi}{dt}.$$ Using chain rule $\frac{d\phi}{dt} = \frac{d\phi}{ds} \frac{ds}{dt}$, and noting that $\frac{ds}{dt} = v_h^\parallel = v_r\sin(\psi_b)$, we have   
$\frac{d\psi_b}{dt} = \frac{v_r}{l_r} \sin(2\psi_b) -v_r\sin(\psi_b)\frac{d\phi}{ds}$. 
For a circular chamber with radius $R$, $\phi(s) = \frac{s}{R}$, and we have 
\begin{equation}
\frac{d\psi_b}{dt} = \frac{v_r}{l_r} \sin(2\psi_b) - \frac{v_r}{R}\sin(\psi_b),  
\label{eq:psit}
\end{equation}
which has an unstable fixed point at $\psi_b =0$ (for $l_r < 2R$), and a stable fixed point, 
\begin{equation}
\label{eq:Stable angle}
\psi_o = \cos^{-1} \left( \frac{l_r}{2R} \right).
\end{equation}
We numerically integrate Eq.~(\ref{eq:psit}) for three different initial rod orientations $\psi_b(0) = 0.01 \psi_o, 0.25 \psi_o$, and $0.5 \psi_o$. The resulting plot in Fig.~\ref{fig:analytic soln}(d) illustrates how the rod's orientation evolves towards the stable point $\psi_0 \simeq 1.19$\,rad in each case. The value of  stable point depends on the ratio of the chamber radius and rod length $R/l_r$ as given in Eq.~(\ref{eq:Stable angle}). A plot of $\psi_o$ versus $R/l_r$ is shown in Fig.~\ref{fig:analytic soln}(e). It can be seen that as $R \rightarrow \infty$, $\psi_o \rightarrow \pi/2$, consistent with the boundary alignment observed near flat boundaries. 

Nonetheless, the measured angle corresponding to experiments in the circular chamber shown in Fig.~\ref{fig:convex_corner}(b) is lower compared with the calculated trend (see Fig.~\ref{fig:analytic soln}(e)). To explain the lower observed value and the trapping–escape dynamics at concave corners, we next develop the model to include the fluctuations observed in the worm’s motion. 

\subsection{Rod Dynamics with Translational and Rotational Diffusion}
\label{sec:simmodel}
As discussed previously~\cite{kudrolli2019burrowing}, {\it L. variegatus} performs peristaltic and undulatory strokes while moving through water and sediment. These strokes contribute to body-orientation, position and body-length fluctuations. In order to capture the effects of the strokes in the SPR model, we incorporate the translation and rotational diffusion in the dynamics. We perform numerical simulations of the SPR model following the framework of Ref.~\cite{Bar2020}. The overdamped Langevin dynamics is implemented using the Euler-Maruyama algorithm to update the rod center and orientation angle, 
\begin{gather}
    \dot{\mathbf{r}}(t) = v_r\mathbf{e}_\parallel + \sqrt{2D_t}\ (\mathbf{e}_\parallel\eta_\parallel(t)  +  \mathbf{e}_\perp\eta_\perp(t)),\,\,\,\, {\rm and}\\ 
    \dot{{\theta}}(t)  = \sqrt{2D_r} \ \xi(t),
\end{gather}
where, $\mathbf{r}(t)$ and ${\theta}(t)$ represent the position vector of rod's centroid and orientation at time $t$, respectively, $D_t$ is the translational diffusion coefficient, which is assumed to be isotropic for simplicity, $D_r$ is the rotational diffusion coefficient, $\eta_\parallel(t)$, $\eta_\perp(t)$ and $\xi(t)$ are white noise processes, $v_r$ is the constant propulsion speed, and $\mathbf{e}_\parallel$ and $\mathbf{e}_\perp$ are the unit vectors parallel and perpendicular to the rod axis, respectively. If the displacement fluctuations due to translational diffusion and rotational diffusion are similar in magnitude~\cite{Baskaran2010}, i.e., $\sqrt{D_tdt} \sim  \frac{l_r}{2}\sqrt{D_rdt}$, then ${D_t}  \approx \frac{1}{4} \, D_r {l_r^2}$. Here, we consider the rod length $l_r$ to be constant, and consider the additional effect of length fluctuations caused by the body strokes in Section~\ref{sec:lenfluc}.

The length and time scales in the simulations are set using the rod length $l_r$ and the time required to translate distance equal to its length while moving with velocity $v_r$, respectively. The boundary conditions are enforced such that if the head end of the rod extends beyond the confinement, the velocity component normal to the boundary, at the head-end, is set to zero to implement a sliding boundary condition. If the tail end of the rod extends beyond the confinement, the time step is rejected and repeated.

\begin{figure}
\centering
\includegraphics[width = 0.8\linewidth]{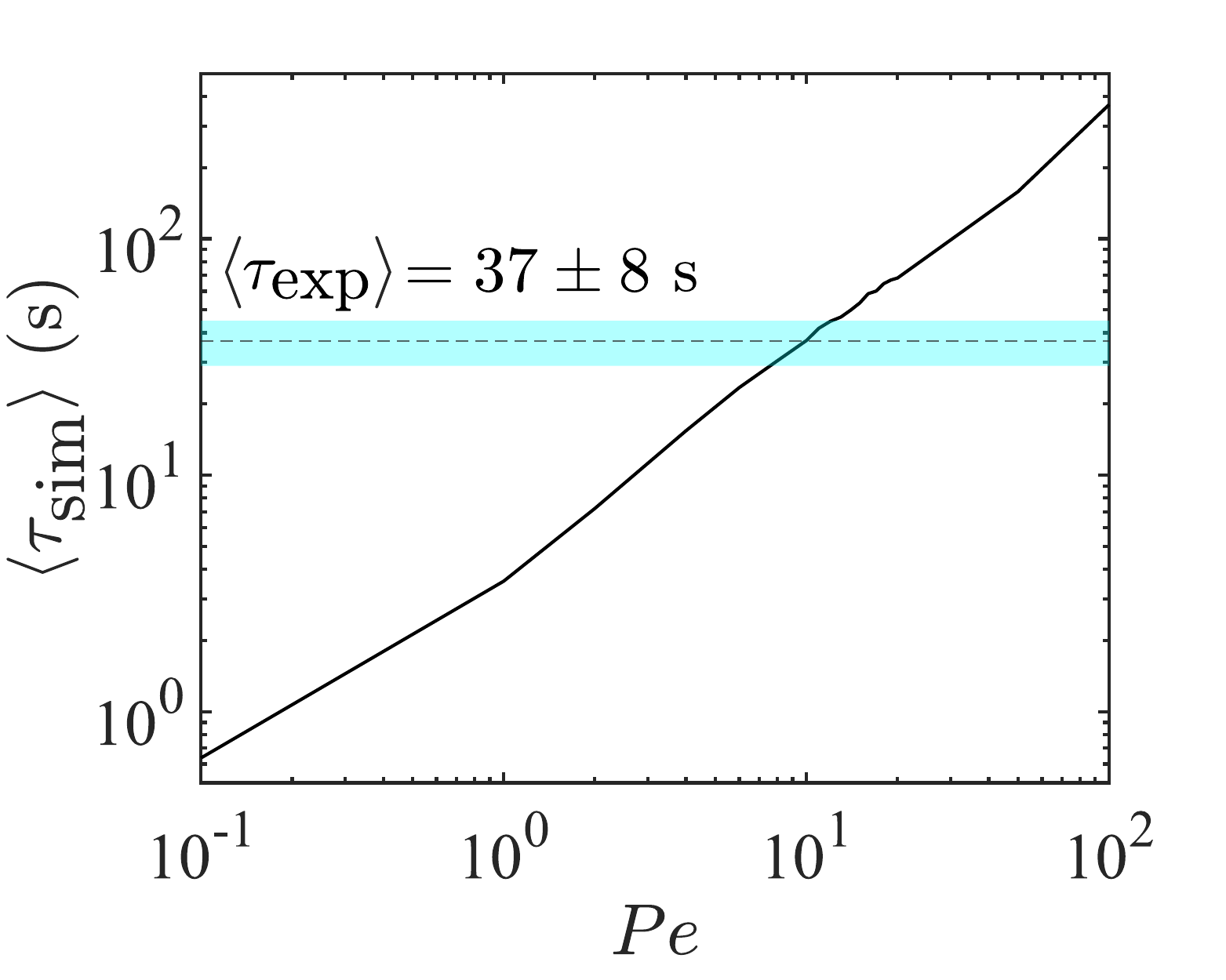} 
 \caption{The mean trapping time of the rod $\langle \tau_\textrm{sim} \rangle$ at a corner of square chamber ($l = 40$ mm) increases with increasing $Pe$. At $Pe = 13$, $\langle \tau_\textrm{sim} \rangle$ matches the trapping time scale observed in experiments within the deviation (mean - dashed line, deviation - shaded region).}
\label{fig:Pe_sweep}
\end{figure}

The relative strength of the forward transport compared to the rotational diffusion 
can be characterized by the rotational P\'eclet number $Pe \sim \frac{v_r}{D_r l_r}$~\cite{Cugliandolo2015}.  The P\'eclet number of the worm using its corresponding parameters can be obtained as $Pe_\textrm{worm} \sim \frac{v_h}{D^{w}_rl_e} \approx 6$, where $l_e\approx15$\,mm is the measured average worm end-to-end length in the experiments. The rotational diffusion coefficient of the worm $D^{w}_r$ is estimated using its tracked trajectory as explained in the SI~\cite{sup-doc}. Because of the differences between a worm and its rigid rod approximation, we can anticipate that $Pe$ for the SPR model may not be the same as the $Pe_\textrm{worm}$. Varying $Pe$, we find that the mean trap time of the rod in a corner $\langle \tau_\textrm{sim} \rangle$ obtained in the SPR simulations increases systematically (see Fig.~\ref{fig:Pe_sweep}). The observed mean trap time observed at the corners $\langle \tau_{\textrm{exp}} \rangle = 37$\,s and its deviation $\pm 8$\,s are also plotted in Fig.~\ref{fig:Pe_sweep}. Using their overlap, we find $Pe \approx 13$, which is reasonably close to $Pe_{\rm worm}$ considering the approximations in the SPR model. 

\begin{figure*}
 \centering
 \includegraphics[width =0.75\textwidth]{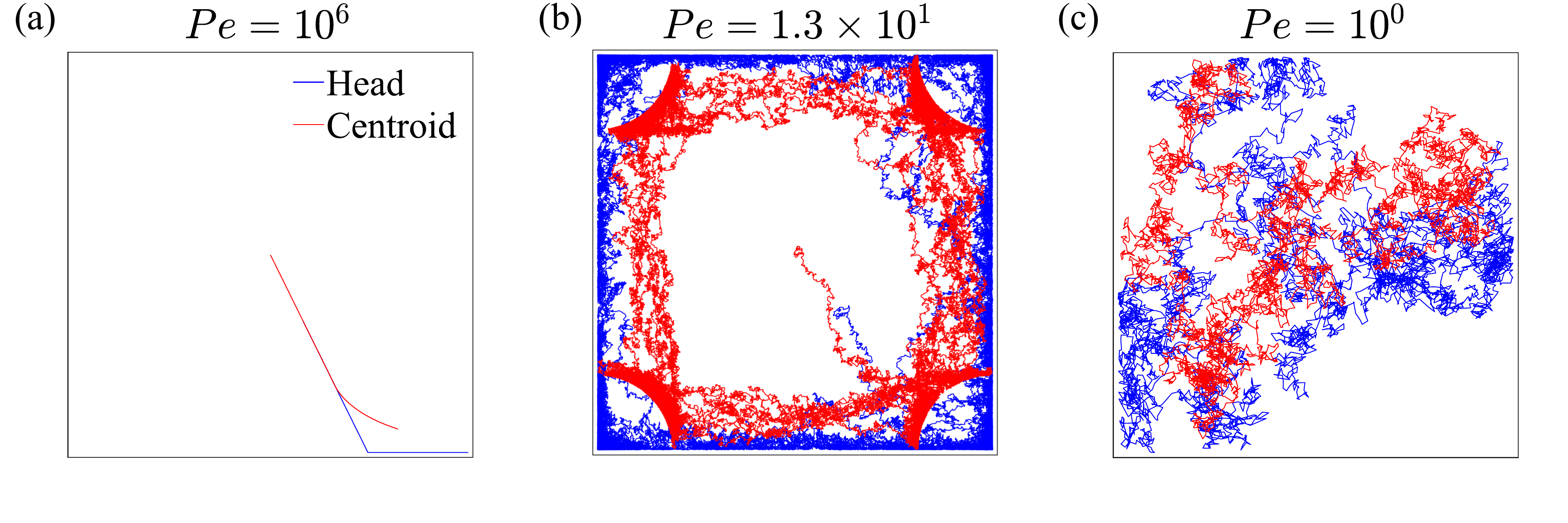}
\caption{The SPR model simulation based trajectories of the rod in a square confinement corresponding to different values of  $Pe$ represents the three dynamical regimes.  (a) At high $Pe$, the rod exhibits boundary following and aligning behavior until it gets permanently trapped in a corner. (b) At intermediate $Pe$, the model captures dynamical trapping at corners along with boundary following and aligning behavior. This regime qualitatively corresponds to the experimental observations with \textit{L. variegatus}.  (c) At low $Pe$, the rod exhibits diffusion motion without showing any boundary aligning and corner trapping features.}
 \label{fig:dynamical_regimes}
\end{figure*}

Representative trajectories are shown in Fig.~\ref{fig:dynamical_regimes}(a-c) with increasing $Pe$. At high $Pe$, the rod moves in a straight line from the center, rotates while moving forward as it comes in contact with the boundary, and then gets trapped indefinitely at the corner. In this limit of the model where $D_r \rightarrow 0$ and $D_t \rightarrow 0$, it approaches the kinematic model as discussed earlier. At $Pe=1$ the rod neither remain near the boundaries nor becomes trapped at the corner . Simulations performed with $Pe=13$ show boundary following, corner trapping and escape reminiscent of the trajectory of {\it L. variegatus} shown in Fig.~\ref{fig:spatial}(a). These results demonstrate that different dynamical regimes can be attained simply by varying $Pe$, including the behavior seen in the experiment. 

\begin{figure*}
\centering
\includegraphics[width = 0.6\textwidth]{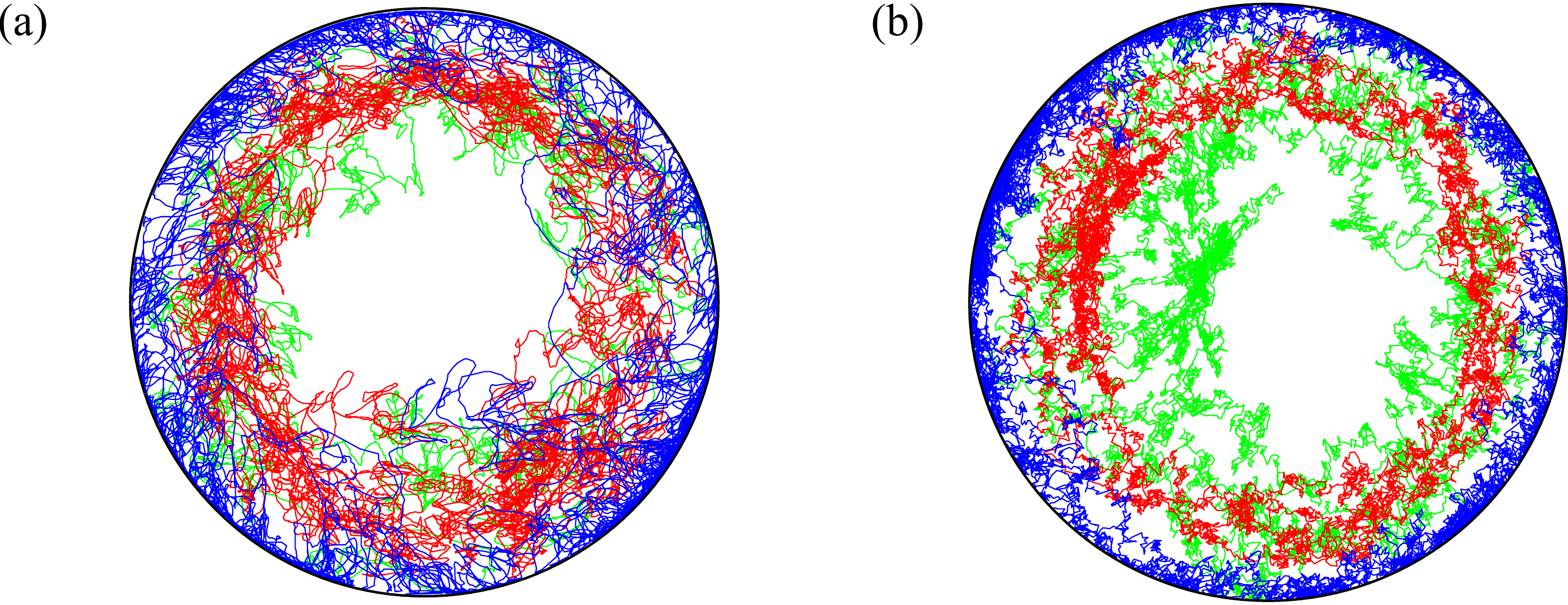} 
 \caption{(a) A trajectory of the worm head (blue), center (red) and tail (green) in a circular chamber over time $T = 10$\,minutes ($R = 2$\,cm; $l_w = 28 \pm 2.3$\,mm). (b) Simulated trajectory of the rod's head tip (blue), center (red) and the tail tip (green), in a circular chamber over time $T = 10$\,minutes ($R = 2$~cm ; $l_r = 15$~mm). }
\label{fig:centroid_circle}
\end{figure*}

We implement the SPR simulations in the circular chamber with $R = 2$~cm. We update the $Pe$ proportionately based on the measured worm speed in the circular chamber $v_h \sim 1.25\textrm{ mm s}^{-1}$ in those examples, yielding $Pe \sim 9.5$. A representative trajectory observed in the experiments and obtained by performing simulations using the resulting parameter $Pe = 9.5$ are shown side by side in  Fig.~\ref{fig:centroid_circle}(a) and Fig.~\ref{fig:centroid_circle}(b), respectively. We further plot the mean boundary orientation in the simulations for varying $R/l_r$ in Fig.~\ref{fig:analytic soln}(e). In the experiment, the mean boundary orientation angle $\psi_b \approx 0.8 \pm 0.2$ rad, while in the simulations, the $\psi_b \approx 0.6 \pm 0.3$ rad, within statistical fluctuations, and 
systematically lower than $\psi_o \simeq 1.19$ rad, from kinematic model for aspect ratio $R/l_r \simeq 1.3$  as shown in Fig.~\ref{fig:analytic soln}(e). These simulations show that while body strokes propel the swimmer forward and help it align with the boundary, they also play a critical role in systematically reducing the boundary alignment angle due to the body fluctuations.

Simulations corresponding to circular, square, and polygonal chambers shown in Fig.~\ref{fig:convex_corner}(b-d) can be found in the SI~\cite{sup-doc}. In the circular chamber, the rod moves along the boundary, occasionally switching direction due to fluctuations in its orientation relative to the wall. In the square chamber, the rod tends to align with the boundary and move toward a corner, where it becomes temporarily trapped for varying durations. In the polygonal chamber, the simulations show that the rod becomes trapped at concave corners and moves into the chamber rather than around the convex corner. Thus, the simulated behavior of the SPR after incorporating diffusion is consistent with the experimental observations in the worms.
 
\subsection{Boundary alignment}
 \begin{figure*}[htb!]
    \centering
    \includegraphics[width=0.8\textwidth]{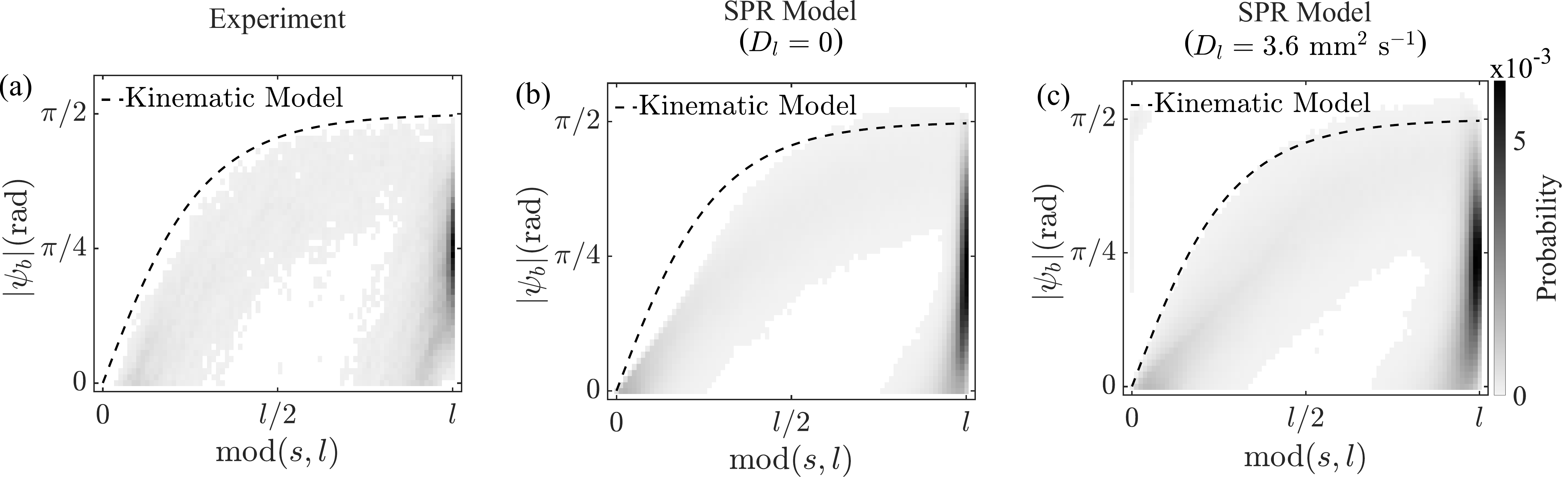}   
    \caption{Comparison of boundary alignment between the experiments and the SPR model simulations. Plot of $|\psi_b|$ versus {mod($s$, $l$)}, where the intensity of the gray-map represents the probability of attaining corresponding configuration (mod($s,l$) ,$\psi_b$) in experiments (a), SPR model with $D_l = 0$ (b) and SPR model with $D_l = 3.6 \textrm{ mm}^2\textrm{ s}^{-1}$ (c). The dashed (black) lines correspond to Eq.~(\ref{eq:psib}) calculated with the kinematic model, for initial condition $s(0) = 0$ mm and $\psi_b(0) = 0.01$ rad} 
    \label{fig:boundaryalignment}
\end{figure*}

To directly compare experimental observations with model predictions of boundary alignment, we take advantage of the square chamber's fourfold rotational symmetry, as well as the symmetry in the worm’s clockwise and counterclockwise motion. The probability of finding a worm's body with a particular $\psi_b$ versus the modulus of its location along the perimeter with respect to the side length, i.e., mod$(s,l)$, is shown in Fig.~\ref{fig:boundaryalignment}(a). We observe two bands which are shaded dark according to the gray map, and can be noted to be a superimposed version of the data shown in Fig.~\ref{fig:boundary}(b). One band corresponds to the worm moving along the boundary, and the other to the worm becoming trapped in the corners. Plotting Eq.~(\ref{eq:psib}) with $s(0) = 0$\,mm, $\psi_b(0) = 0.01$\,rad, and $l_r = 15$ mm with dashed line in Fig.~\ref{fig:boundaryalignment}(a), we find that it bounds the scatter of the highest density data points as the worm moves along the boundary from one corner to an adjacent one. 
The corresponding $\psi_b$ distributions obtained with the SPR model are shown in Fig.~\ref{fig:boundaryalignment}(b), where Eq.~(\ref{eq:psib}) is also plotted using the same initial conditions as in the experiments. Comparing the numerically simulated and experimentally measured distributions, we observe overall similar pattern of high probability bands. 

\subsection{Corner trapping and escape}

\begin{figure}
    \centering
    \includegraphics[width=\linewidth]{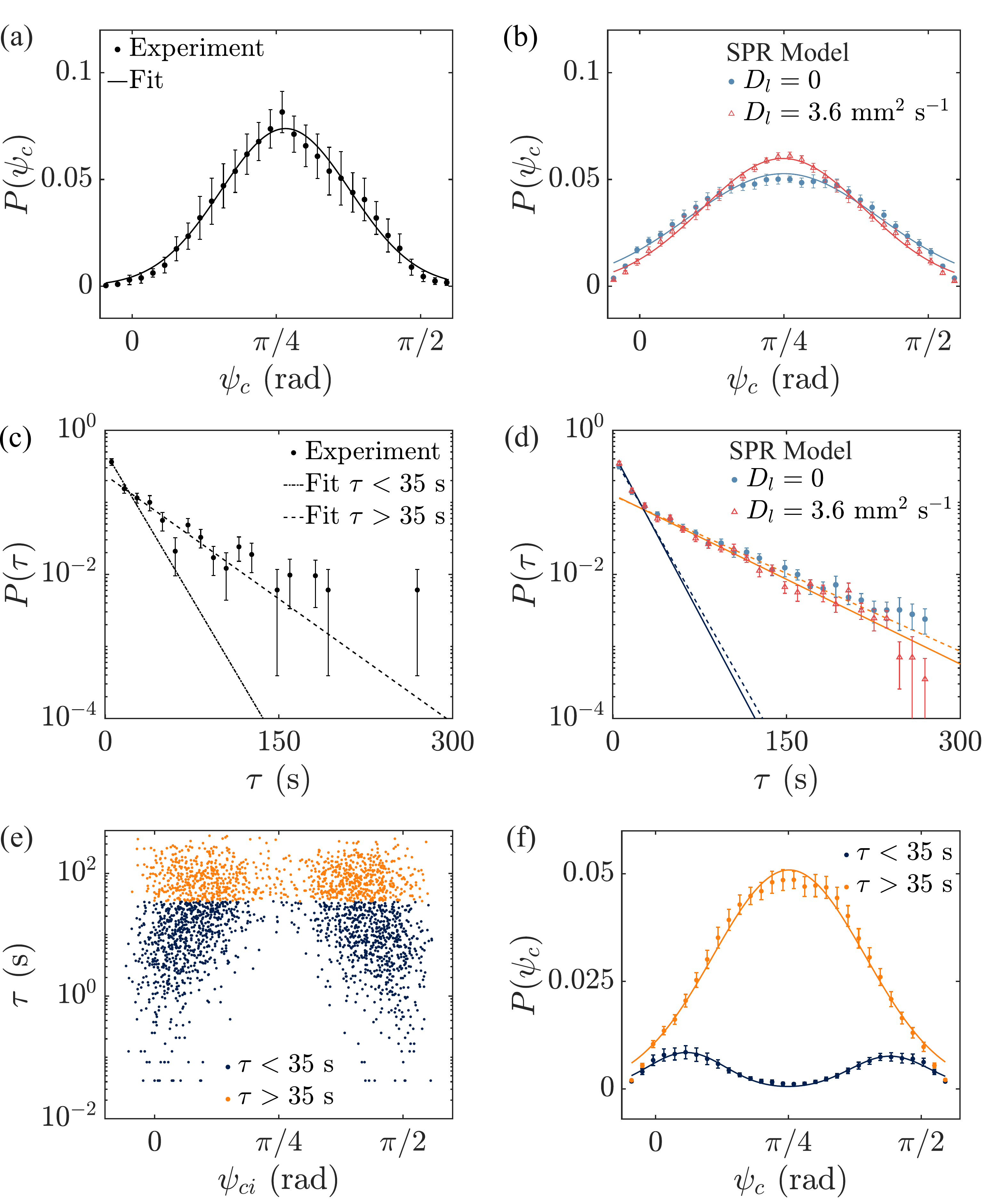}
    \caption{(a) $P(\psi_c)$ of a worm when trapped at a corner and a Gaussian distribution fit with mean $\psi_c = 0.84$\, rad and standard deviation $0.34$\,rad. The bars errors correspond to RMS deviations from the mean. (b) The corresponding distributions for the SPR model without and with fluctuations, and Gaussian distribution fits with mean $0.78$~rad and standard deviation $0.52$~rad, and mean $0.78$~rad and standard deviation $0.44$~rad, respectively. (c) $P(\tau)$ of a worm at a corner is best described using two exponential forms with means $\langle \tau_a \rangle = 16$~s, $\langle \tau_b \rangle = 38$~s while $\tau < 35$ s~and $\tau > 35$~s, respectively. (d) The corresponding distributions and fit for the SPR model with mean trap time $\langle \tau_a \rangle = 15$~s, $\langle \tau_b \rangle = 58$~s, and with fluctuating length with mean trap time $\langle \tau_a \rangle = 14$~s and $\langle \tau_b \rangle = 56$~s. (e) The scatter plot of trapping time $\tau$ and incidence angle $|\psi_{ci}|$ at a corner. (f) $P(\psi_c)$ corresponding to the two groups. The probability corresponding to $\psi_c = \pi/4$ for the $\tau < 35$~s distribution is negligible, indicating that the rod escapes rapidly without reorienting along the angle bisector.}
    \label{fig:exp_vs_sim}
\end{figure}

We now examine corner trapping by plotting the probability distribution $P(\psi_c)$ when the worm's head is in the corner region. Here, $\psi_c$ is the orientation of the worm at the corner, defined as the angle that the line joining the centroid to head makes with the outward normal to the successive boundary. At a corner, the boundary is assumed to continue along with increasing $s$. As shown in Fig.~\ref{fig:exp_vs_sim}(a), the body strokes produce a broad distribution centered around the perpendicular bisector angle. The finite probability near $\psi_c \leq 0$ and $ \geq \pi/2$ radians corresponds to the worm being aligned parallel to the boundary and to escape events from the corner. We also plot $P(\psi_c)$ from simulations where the rod’s leading end is confined at the corners in Fig.~\ref{fig:exp_vs_sim}(b) and find that it is broadly distributed. Thus, the worm’s orientation at corners is governed by its intrinsic forward motion and body strokes, along with interactions with the boundaries, at the corner. Gaussian fits to the distributions yield values $\mu_\textrm{exp} = 0.84$\,rad, $\mu_\textrm{sim} = 0.78$\,rad, and standard deviations $\sigma_\textrm{exp} = 0.34$\,rad, $\sigma_\textrm{sim} = 0.52$\,rad, systematically broader in the SPR model. As we will discuss later, a narrower distribution can be observed in simulations by including the effect of the length fluctuations. 

To characterize the trapping duration, we measure the corner trapping time $\tau$ and plot its distribution $P(\tau)$ along with the root-mean-square (RMS) fluctuations across trials, as shown for the experiments in Fig.~\ref{fig:exp_vs_sim}(c)  and for the SPR simulations in Fig.~\ref{fig:exp_vs_sim}(d). While, the mean of the overall trapping time distribution was used to obtain $Pe$ in the simulations, it is not sufficient to fully characterize the observed distributions. Noting the semi-log scales, we observe that $P(\tau)$ decreases rapidly at small $\tau$, and then decreases slowly with increasing $\tau$, in the experiments and in the simulations.  Exponential fits to $P(\psi_c)$ are also plotted over the range $\tau < 35$\,s, and $\tau \geq 35$\,s to guide the eye. These combined fits provide a better description of the data than a single exponential, which cannot capture the observed distribution.

\subsection{Length fluctuations} \label{sec:lenfluc}
We have so far assumed a constant $l_r$  corresponding to the average worm end-to-end length $l_e$. In reality $l_r$ fluctuations can be caused due to the body strokes observed in the \textit{L. variegatus} ~\cite{kudrolli2019burrowing}. This feature can be incorporated into the simulations by allowing $l_r$ to vary stochastically, governed by an appropriate diffusion coefficient $D_l$ while $l_r$ is constrained within upper and lower bounds $l_\textrm{max}$ and $l_\textrm{min}$, respectively, due to the finite worm length. The dynamics of $l_r$ are modeled as $\dot{l_r}(t) = \sqrt{2D_l}\,\xi(t)$, where $\xi(t)$ is a white noise process. At each time step, based on $l_r(t)$, $\dot{\mathbf{r}}(t)$ and $\dot{{\theta}}(t)$ are updated accordingly. The boundary conditions are implemented as described earlier in Section~\ref{sec:simmodel}.

We simulate the SPR model where the rod length varies over time, using $D_l =  3.6\textrm{ mm}^2\textrm{ s}^{-1}$ with $l_\textrm{max} = 1.5 \ l_r$ and $l_\textrm{min} = 0.5 \ l_r$ obtained from the mean-squared worm end-to-end length fluctuations observed in the experiments (see SI document). The resulting $\psi_b$ distributions are shown in Fig.~\ref{fig:boundaryalignment}(c). While, these distributions are qualitatively similar, deviations can be noted in comparison with the kinematic model. In particular, the SPR model including the length fluctuations produces distributions which more closely bound the kinematic model, closely resembling those recorded in the experiments shown in Fig.~\ref{fig:boundaryalignment}(a).  Thus, body fluctuations reduce the degree of boundary alignment and increase the angle between the worm’s body and the boundary as the head moves along it; nonetheless, the overall boundary-aligning dynamics remain intact.

We obtain $P(\psi_c)$ with the SPR model while the rod length varies ($D_l = 3.6\textrm{ mm}^2\textrm{ s}^{-1}$) and plot it in Fig.~\ref{fig:exp_vs_sim}(b). Gaussian fits to the distributions yield values $\mu_\textrm{sim} = 0.78$\,rad, and standard deviations $\sigma_\textrm{sim} = 0.44$\,rad, closer to those observed in the experiments. Moreover, we observe that $P(\tau)$ more or less overlap when $D_l=0$ and $D_l = 3.6$\,mm$^2$s$^{-1}$. Thus, while the fluctuating end-to-end length leads to a narrowing of $P(\psi_c)$, it has little effect on the mean escape time itself.

\subsection{Effect of entry angles on trapping time}

We investigate the relationship between the initial corner entry angle  $\psi_{ci}$ and trapping time $\tau$ in the SPR model assuming constant rod length to assess the correlation between trapping time and orientation distributions. Figure~\ref{fig:exp_vs_sim}(e) shows a semi-log scatter plot of $\tau$ versus $\psi_{ci}$. A symmetric distribution of points around $\psi_{ci} = \pi/4$\,rad is observed, indicating complementary behavior when encountering a corner from either the clockwise direction ($\psi_{ci} = \pi/2$\,rad) or the anti-clockwise direction ($\psi_{ci} = 0$). The high density of points near  $\psi_{ci} = 0$ and $ \pi/2$ radians in Fig.~\ref{fig:exp_vs_sim}(e) indicates that the rod tend to align along with the chamber sides before becoming trapped at a corner. We observe that most of the events with initial orientation in vicinity of $\psi_{ci} = \pi/4$\,rad lead to a higher trapping time, while at lower trapping time, the majority density of points is localized around $\psi_{ci} = 0,\pi/2$\,rad. 

We divide the trapping events into two groups: short-duration $(\tau < 35\textrm{ s})$, and long-duration ($\tau \geq 35\textrm{ s})$, and plot the corresponding $P(\psi_c)$ distributions separately in Fig.~\ref{fig:exp_vs_sim}(f). For the trapping events with $\tau < 35 \textrm{ s}$, the distribution is well described by a two-component normal fit with means $\mu_{1} = 0.18 \textrm{ rad}$ and $\mu_{2} = 1.39\textrm{ rad}$, and standard deviations $\sigma_1 = 0.23\textrm{ rad}$ and $ \sigma_2 = 0.23 \textrm{ rad}$. The two distributions arise from the symmetry of the confinement and correspond to events where the rod approaches a corner nearly parallel to one of the chamber edges and escapes quickly. As clearly seen in Fig.~\ref{fig:exp_vs_sim}(f), the probability of $\psi_{c} = \pi/4$\,rad is nearly zero, implying that the rod does not get trapped along the angle bisector at the corner. When the rod escapes from a corner it orients to angles $\psi_c < 0$ or $ \psi_c > \pi/2 $\,rad. The orientation probability associated with the population with high $\tau$ is well described by a normal distribution with mean $\mu \approx 0.79$\,rad and $\sigma \approx 0.45$\,rad. These events correspond to instances where the rod aligns along the corner’s angle bisector. The peak of this distribution is higher than the individual peaks in the short-$\tau$ population, reflecting cumulative contributions from cases where the rod approaches the corner with initial orientations in the range $0 < \psi_{ci} < \pi/2$.  In this regime, the rod effectively loses memory of its initial orientation during the trapping process.

From this analysis, we conclude that the shape of the distributions shown in Fig.~\ref{fig:exp_vs_sim}(b) arise from the combined effect of two distinct populations: one consisting of rods that enter and exit the corner almost immediately, and another comprising of rods that remain trapped for variable durations, governed by a stochastic process, while their orientation fluctuates around the perpendicular bisector of the corner. Thus, even though the $\tau$-distributions deviate from exponential form from a Poisson process, the deviations primarily arise due to the influence of the entry angle, rather than any attraction interaction that would lead to a non-Poisson process. 

More generally, we find that the boundary alignment, corner trapping, and escape behaviors observed in these California blackworms emerge naturally from the intrinsic dynamics of the SPR model, without requiring any explicit attractive forces toward walls or corners.

\section{Conclusions}
In this study, we examined the behavior of {\it Lumbriculus variegatus} within confined geometries of varying curvature, focusing on polygonal chambers with concave and convex corners, and compared the results with those obtained in circular chambers. Across all geometries, the worm consistently exhibited a higher probability of occupying regions near the boundaries than the central area. However, the polygonal chambers revealed richer and more nuanced interactions between the worm and the confinement, particularly at corners where the nature of boundary curvature becomes more prominent.

Leveraging full-body tracking of the worm, we analyzed how boundary alignment and trapping emerge from the interplay between body strokes and physical confinement. We observed that worms tend to localize near concave corners but not at convex corners, where they continue along their trajectory without following the boundary. Moreover, worms were seen to spontaneously leave both the boundaries and corners, suggesting that localization arises from the intrinsic dynamics of their motion rather than any shelter-seeking behavior or thigmotaxis. These findings point to a primarily dynamical origin of confinement-induced localization.

To better understand these dynamics, we began by constructing a kinematic model that approximates the worm as a rigid rod, with its mean velocity aligned along its axis. By incorporating a frictionless sliding contact condition at the boundary, we analytically determined how the worm’s orientation evolves as a function of its trajectory and speed along the boundary. This model provides a quantitative description of the mean boundary-alignment behavior observed in experiments with both circular and polygonal chambers.

Building upon the kinematic model, we developed a self-propelled rigid rod (SPR) model that incorporates translational and rotational diffusion arising from the stochastic nature of body strokes. We further extended the model to include fluctuations in the worm’s end-to-end length, which improved the agreement with orientation distributions during corner trapping. Simulations using this model successfully reproduced key experimental features, including boundary alignment, corner trapping, and escape dynamics. A key parameter emerging from the model is the P\'eclet number, which quantifies the ratio of directed motion to rotational diffusion driven by the worm's undulatory and peristaltic body strokes. This dimensionless number governs both the extent of boundary alignment and the characteristic timescales of trapping, establishing a direct link between locomotion dynamics and spatial behavior.

The SPR simulations further reveal that stochastic fluctuations are critical for enabling escape from concave-corner traps. While length fluctuations improve orientation statistics, they have minimal impact on the distribution of trapping durations. Our numerical analysis shows that corner escape time distributions are inherently non-exponential, a result of the worm’s incident orientation upon entering a corner. These distributions arise from a superposition of two distinct behavioral modes: a population of worms that briefly contact the corner and escape almost immediately, and another that reorient along the corner perpendicular bisector and remain trapped for variable durations. The later population escapes the corners governed by stochastic fluctuations in orientation around the perpendicular bisector of the corner. Thus, while the escape time distributions are non-exponential and the orientation angle distributions are non-Gaussian, they arise due to the initial entry conditions into the corners, and not due to any attraction or such feature of the worm. 

While active polymer models \cite{fazelzadeh2023active,martin2024tangentially,deblais2020phase,ozkan2021collective}, have been successfully applied to describe worm-like locomotion in other contexts, they are not strictly necessary for blackworms in our system. This is because boundary interactions are dominated by the worm’s head, which is typically the only part in direct contact with the boundary. As a result, the reorientation dynamics can be captured by a simplified rod model that accounts for the head position and incorporates the mean length and its fluctuations as effective degrees of freedom. This approximation is sufficient to reproduce the observed behavior without invoking the full complexity of an active polymer representation.

As noted in the introduction, previous studies have reported the accumulation of organisms near boundaries and corners. However, these findings have primarily involved micro-swimmers that operate in low Reynolds number regimes and interact with boundaries through long-range hydrodynamic effects, which differ fundamentally from the contact-interaction dynamics of our system. Observation of bacterial accumulation near surfaces in a channel did not track body shape or investigate corner trapping and escape~\cite{Li2009microswimmers}. Similarly, Galajda, {\it et al.}\cite{galajda2007wall} showed that funnel-shaped boundaries could concentrate run-and-tumble bacteria, but without tracking body configurations or detailed trajectories. Other studies, such as Th\'ery et al.~\cite{Thery2021Rebound}, attributed corner accumulation in model systems to specular reflection mechanisms, which are distinct from those at play in our study. Work on bacterial motion in micro-chambers with and without scattering sites added, show presence or absence of corner accumulation~\cite{frangipane2019invariance}, which again is distinct from the mechanism of corner trapping and escape elucidated by our study.  Theoretical models~\cite{solon2015active,khatami2016active} have also contrasted aggregation in the presence and absence of persistent motion  and physical regimes that are not directly comparable to the boundary interactions we observe in {\it L. variegatus}. 

Together, our experimental observations and modeling demonstrate that localization near boundaries and corners can arise entirely from contact-mediated dynamics in a regime where long-range hydrodynamic interactions are absent. The self-propelled, polar nature of the worm’s motion, coupled with limited rotational diffusion due to its elongated shape, is sufficient to explain the observed behaviors. This occurs without the need for  behavioral mechanisms such as thigmotaxis or contact seeking behavior. While alternative behavioral mechanisms, such as thigmotaxis or active corner-seeking, cannot be fully excluded, our results indicate that passive geometric constraints and steric interactions are sufficient to explain the observed corner trapping within the limits of our experimental resolution. Overall, our work underscores the importance of boundary geometry in shaping the motion of confined undulatory organisms and offers a minimal yet predictive framework to understand how simple physical rules can give rise to complex behaviors in biological systems under confinement.

\section*{Data Availability Statement}
The data supporting this article has been included as a part of the main article and the Supplementary Information.

\section*{Conflicts of interest}
There are no conflicts of interest to declare.

\begin{acknowledgments}
We thank Animesh Biswas, Vishal Patil, and Simon Bissitt for discussions. We acknowledge the support of the U.S. NSF grant CBET-1805398 and 2401729, and thank Professor Alex Petroff for allowing access and guidance to the Laser-cutter facility in his laboratory. 
\end{acknowledgments}

\bibliographystyle{unsrt}

\newpage

\appendix

\section{Movies}
List of movies and captions.

\begin{itemize}
    \item Movie S1: A California blackworm moving in a circular chamber is observed to turn and follow the boundary. The frame rate is  $1$ fps and total time is $450$ s.  ($R=20$\,mm; $l_w = 18 \pm 1.50 $\,mm). The movie is played at $10$ fps.
    \item Movie S2: A California blackworm moving in a square chamber is observed to follow the boundary and get trapped at the corners before escaping over varying internal of time. The frame rate is $1$ fps and total time is $600$ s. ($l=40$\,mm; $l_w = 22 \pm 2$\,mm). The movie is played at $10$ fps.
    \item Movie S3: A California blackworm moving in a polygonal chamber with concave and convex corners. The worm is observed to follow the boundary but not across the convex corner. It is observed to get trapped at concave corners as also seen in the square chambers. The frame rate is $1$ fps and total time is $600$ s. ($l=40$\,mm; $l_w = 15 \pm 1.8$\,mm). The movie is played at $10$ fps.
    
    \item Movie S4: An animation of the simulated self- propelled rod (SPR) model in a circular chamber. The frame rate is $4$ fps and total time is $600$ s.($R=20$\,mm; $l_r = 15$\,mm), The animation is played at frame rate of $40$ fps.

    \item Movie S5: An animation of the simulated SPR model in a square chamber. The frame rate is $4$ fps and total time is $600$ s.($l=40$\,mm; $l_r = 15$\,mm), The animation is played at frame rate of $40$ fps.
    
    \item Movie S6:  An animation of the simulated SPR model in a polygon chamber. The frame rate is $4$ fps and total time is $600$ s.($l=40$\,mm; $l_r = 15$\,mm), The animation is played at frame rate of $40$ fps.
 
\end{itemize}

\section{Image Processing}
\begin{figure*}
 \centering
 \includegraphics[width = 0.95\textwidth]{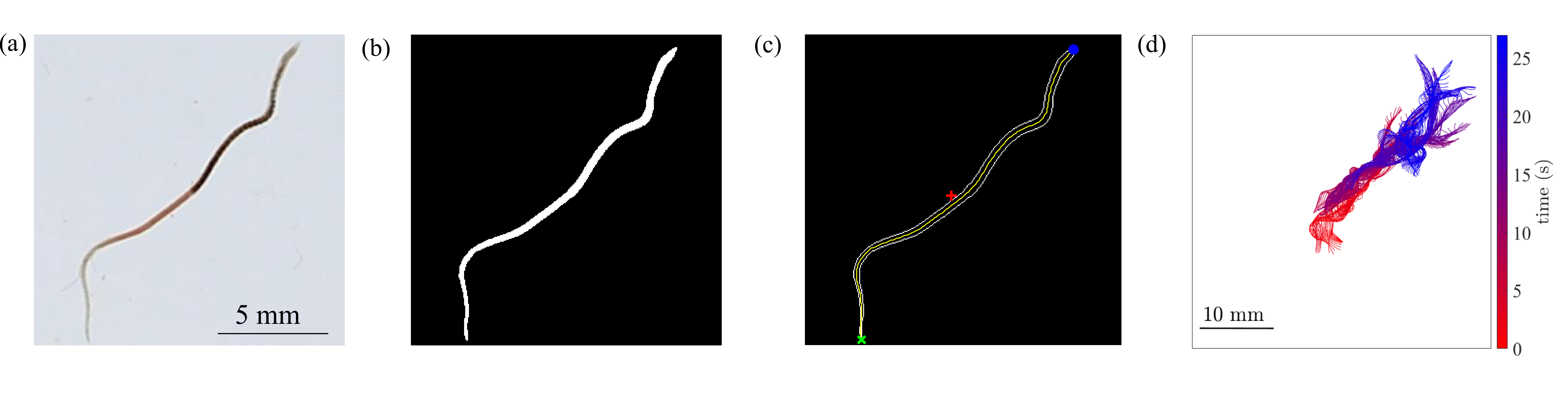}
  \caption{(a) A closeup image of a California blackworm. (b) The complementary binarized image . (c) Tracked body shape (solid yellow curve) of the worm with head (blue o), centroid (red +) and tail (green x) coordinates. (d) Superimposed snapshots of a worm moving diagonally across a square chamber at 12 frames per second or 83.3\,ms time intervals. The color bar represents time.} \label{fig:image}
\end{figure*}

The system is imaged from above with a DSLR camera Canon Rebel T5i with a 18 megapixel CMOS sensor while the worms are back-lit with a uniform LED array.  After the worm acclimatizes over a few minutes, the motion of the worm is recorded over approximately 30\,minutes at 24\,fps, and the worm is returned back to the holding tank. A representative cropped image is shown in Fig.~\ref{fig:image}(a). The images are binarized by using a suitable threshold  and its complement image is shown in Fig.~\ref{fig:image}(b) and the entire shape of the worm is tracked using Image Acquisition Toolbox functions implemented in MATLAB. The tracked worm body position superimposed on an image of the worm are shown in Fig.~\ref{fig:image}(c), and several superimposed snapshots of the worm while moving inside the chamber is shown in Fig.~\ref{fig:image}(d). We obtain the worm locomotion speed, its orientation, the stroke amplitudes, along with the position of the head and tail from the tracked shape, and use them for further analysis.

The recorded images are pre-processed before they can be used for extracting data. We use the following steps for pre-processing:
\begin{enumerate}
\item \textbf{Conversion of an image from RGB to gray scale.} The image captured by the camera is in RGB scale. We convert the image to gray scale. Under the applied lighting conditions, the worm appears dark in contrast to the bright background.

\item \textbf{Subtracting the background from the image.} To remove the background, we average over all the images to obtain a background image. Each frame is then subtracted from this background image. The resulting difference image has the worm body as bright pixels against a dark background. This results in an image where the worm's body is represented as bright pixels against a dark background.

\item \textbf{Binarizing the image.} The gray scale images are binarized using an appropriate threshold. We set the threshold value such that the worm's body pixel value equals 1.

\item \textbf{Cleaning the image.} The noise in the binarized image is removed by determining all the connected objects in the image and choosing the object having the maximum number of pixels which corresponds to the worm body.

\item \textbf{Skeletonizing the image.} We then determine the skeleton of the image. The skeleton image is a 2D curved line representing the body shape of the worm. 
\end{enumerate}

The position coordinates for the head, centroid and tail of the worm can be tracked from the processed image. Using inbuilt MATLAB operations, we can directly extract the position of the centroid of any segment (in this case, the worm body) present in the image. 
To track head and tail, we extract and save the endpoints from the images and connect them using the Hungarian tracking algorithm.
To track the full body shape of the worm we use the skeleton image and sort its pixel from head to tail using the tracked head and tail coordinates.

\section{Contour length of the worms}
We measure the mean and standard deviation of the contour length of worm from 10 trials of the experiment. The length is sampled over the total time $T \approx 30$, min with time interval $\Delta t = 1$ second. Overall, the mean length and fluctuation of the worms is $\l_w = 20 \pm 3$ mm.
\begin{center}
    \begin{tabular}{|c|c|c|}
    \hline
    Experiment No. & \ Mean length & \ Standard deviation \\
                   &$l_w$ (mm)  & (mm) \\
    \hline
     1 & 17.2 & 1.7 \\
     2 & 19.8 & 2.2 \\
     3 & 20.2 & 1.8 \\
     4 & 23.3 & 2.2 \\
     5 & 17.0 & 1.8 \\
     6 & 19.9 & 2.0 \\
     7 & 22.9 & 4.4 \\
     8 & 17.9 & 2.3 \\
     9 & 19.9 & 2.2 \\
    10 & 20.3 & 2.4 \\
    \hline
    \end{tabular}
\end{center}
\section{Locomotion speed of the worms}
Worm speed is evaluated in the central region away from the boundary, in the boundary region, and at the corners. This analysis includes all relevant events from 10 independent 30-minute trials. Average speeds are computed from position data sampled over a time interval of $2.5$\,seconds, selected to smooth out fluctuations over two peristaltic body strokes.
The measured speeds are as follows.\\
\\
Centroid speed:
\begin{align*}
    \textrm{Center},\, & v_{c} = 0.95  \pm 0.49 \ {\rm mm\,s}^{-1} \\
    \textrm{Boundary},\, & v_{c} = 0.82 \pm 0.57 \ {\rm mm\,s}^{-1}\\
    \textrm{Corner},\, & v_{c} = 0.47 \pm 0.15 \ {\rm mm\,s}^{-1}
\end{align*}
Head speed:
\begin{align*}
    \textrm{Center},\, & v_{h} = 1.72 \pm 0.65 \ {\rm mm\,s}^{-1}\\ 
    \textrm{Boundary},\, & v_{h} = 1.33 \pm 0.70 \ {\rm mm\,s}^{-1}\\
    \textrm{Corner},\, & v_{h} = 0.26 \pm 0.07 \ {\rm mm\,s}^{-1}
\end{align*}
Thus, we observe that the velocity of both the centroid and the head while confined in the corner is significantly lower than when it is away from the boundaries due to the collisions with the boundaries. Further, it can be noted from the speeds that the head is more motile than the centroid in each region, except in the concave corner. 

\section{Measurements of Diffusion constants}

\begin{figure}
\centering
\label{Diffusion constant}
\includegraphics[width = 0.7\linewidth]{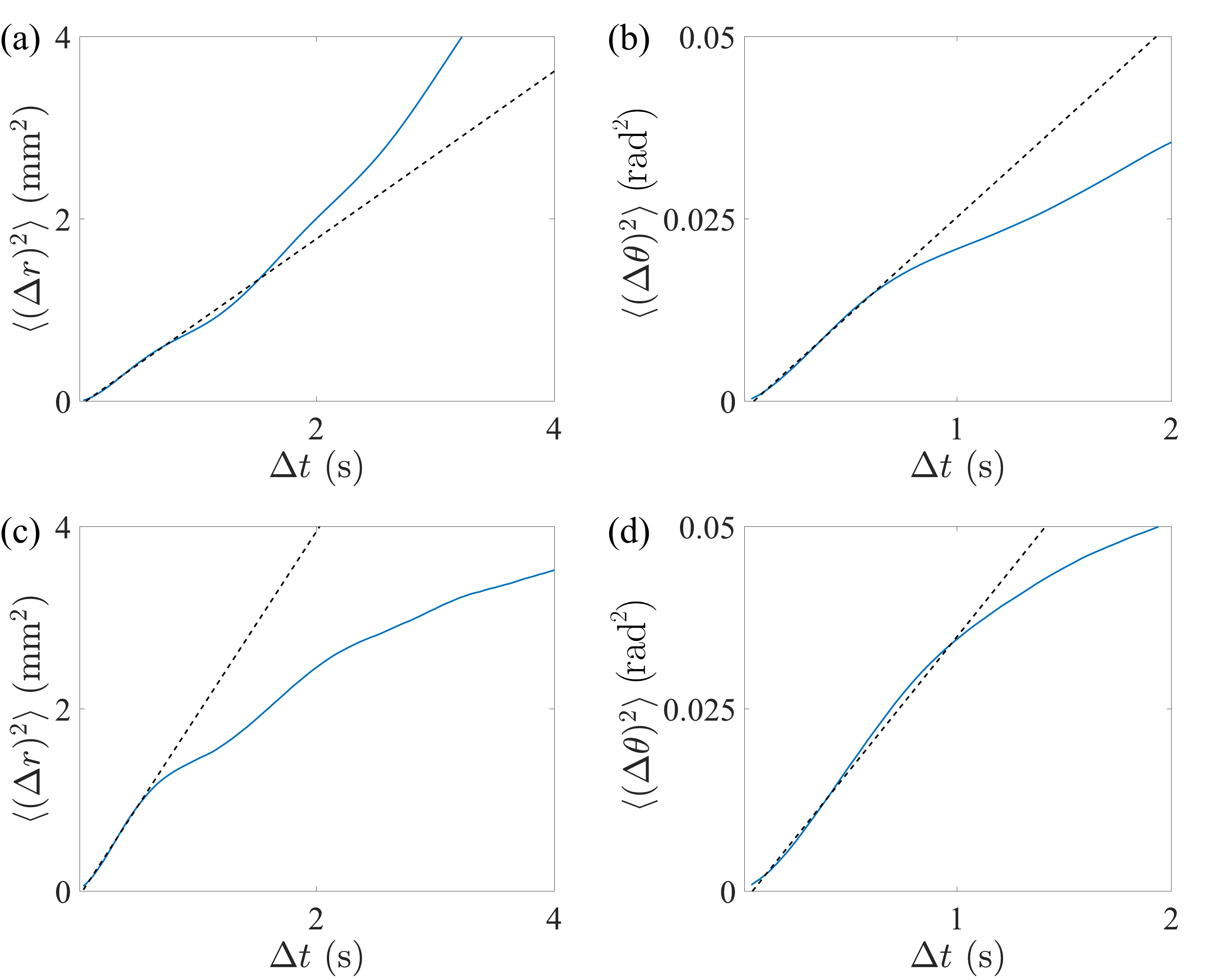}
 \caption{(a) The mean square displacement versus time interval in the circular chamber ($D^w_t = 0.22$ \textrm{mm}$^2$ \textrm{s}$^{-1}$). (b) The mean square displacement versus time interval in the circular chamber ($D^w_r = 0.014 \ \textrm{s}^{-1}$). 
 (c) The mean square displacement versus time interval in the square chamber ($D^w_t = 0.5 \ \textrm{mm}^2 \textrm{s}^{-1}$).  (d) The mean square angular displacement versus time interval in the square chamber ($D^w_r = 0.018 \ \textrm{s}^{-1}$). The black dashed line corresponds to the time interval $\Delta t = 0.5$ s} over which the growth is linear.
\label{fig:Diffusion}
\end{figure}

We obtain the translational diffusion constant $D^w_t$ from the mean square displacement of the worm's center as a function of time interval $\Delta t$, given by $\langle (\Delta r)^2 \rangle = 4 D^w_t \Delta t$. The worm's center is obtained by averaging the worm's head and tail coordinates and hence is different from the worm's body centroid. The rotational diffusion constant $D^w_r$ is similarly determined from the mean square angular displacement, given by $\langle (\Delta \theta)^2 \rangle = 2 D^w_r \Delta t$. Here $\theta$ is the angle subtended by the line joining tail to head with the horizontal. Plots of  $\langle (\Delta r)^2 \rangle$ and $\langle (\Delta \theta)^2 \rangle$ as a function of time interval $\Delta t$ in the circular chamber are shown in Fig.~\ref{fig:Diffusion}(a) and Fig.~\ref{fig:Diffusion}(b), respectively. Similarly, Fig.~\ref{fig:Diffusion}(c) and Fig.~\ref{fig:Diffusion}(d) show the corresponding plots in the square chamber with side length $l =4$\,cm. The time interval $\Delta t = 0.5$~s, over which the mean square displacements grow linearly, is used to fit the data and obtain the two diffusion constants.

\begin{figure}
\centering
\includegraphics[width = 0.35\linewidth]{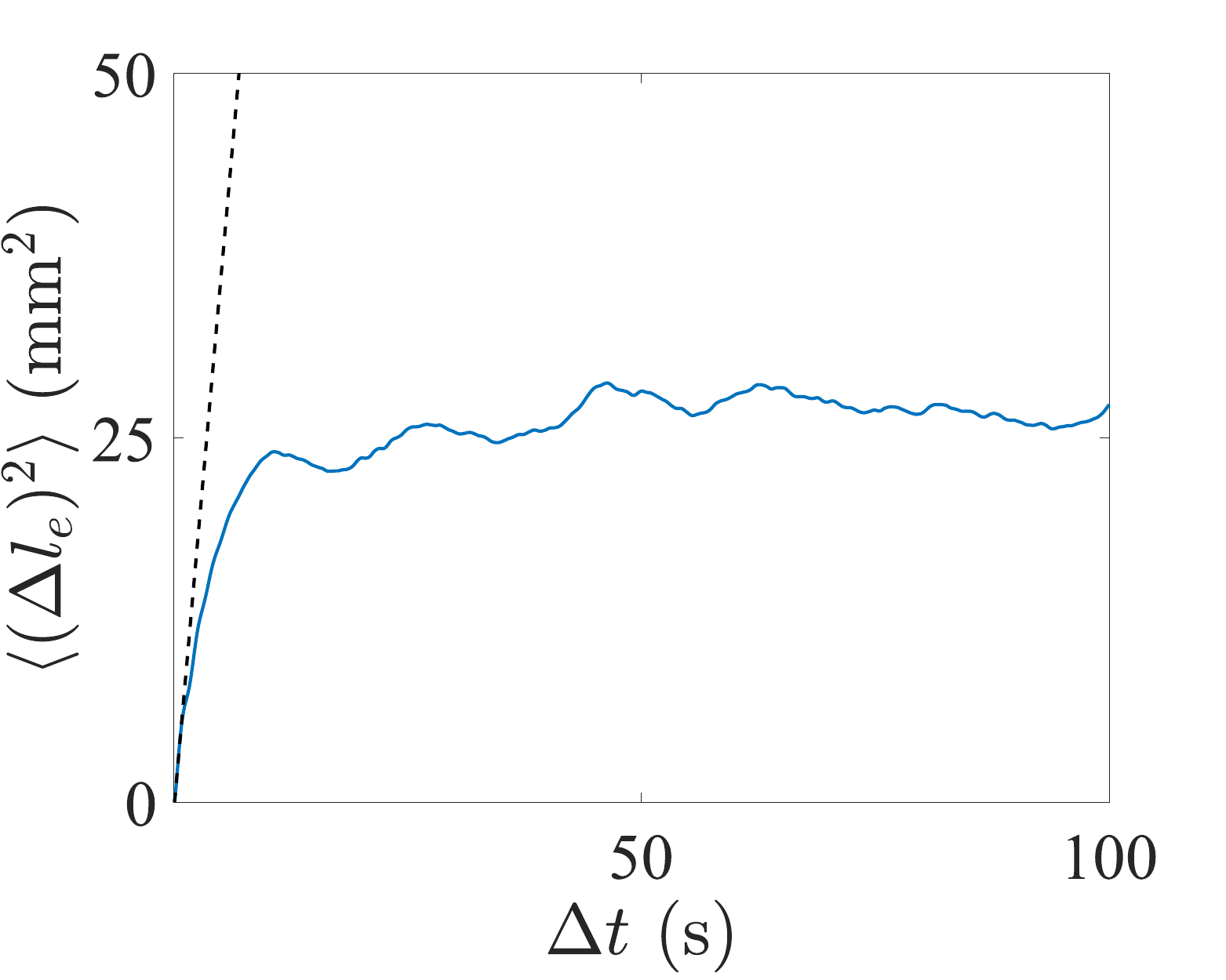}
 \caption{The mean square fluctuation of the worm end-to-end distance as a function of time. ($D^w_l = 3.6$mm$^2$s$^{-1}$). The black dashed line represents the linear fit over time interval $\Delta t = 0.5$ s. The growth over long times is limited by the worm length and its elongation due to the peristaltic stroke. }
 \label{fig:Length fluctuations}
\end{figure}

Further, the diffusion constant $D^w_l$ corresponding to the end-to-end distance length $l_e$ is obtained from the MSF of the worm end-to-end length $\langle (\Delta l_e)^2 \rangle = 2 D^w_l \Delta t$ (see Fig.~\ref{fig:Length fluctuations}). Guided by these measurements, we assume $D_l = D^w_l = 3.6\,$mm$^2$s$^{-1}$ in performing simulations with the SPR model to include end-to-end length variations. The saturation of the MSF of the worm end-to-end length at long times leads to an estimate of maximum deviation in worm-length $\langle \Delta l_e \rangle_\textrm{max} \approx 5$ mm. Thus, we set $l_e^{\rm min} =10$\,mm and $l_e^{\rm max} = 20$\,mm 

\end{document}